\documentclass[aps,prb,twocolumn,superscriptaddress,nobibnotes,amsmath,amssymb,reprint]{revtex4-2}

\usepackage{graphicx}
\usepackage{xcolor}
\usepackage[normalem]{ulem}
\usepackage{natbib}

\newcommand{\twoprod}[8]{\left(s_{#1 #2}^{#3 #4}\right)^{*}s_{#5 #6}^{#7 #8}}
\newcommand{\twoprodpr}[8]{s_{#1 #2}^{#3 #4}s_{#5 #6}^{#7 #8}}

\newcommand{\dnmrt}[0]{1+R_1R_2-2\sqrt{R_1R_2}\left|t_{ee}\right|\cos\Phi}

\begin{document}
\title{Efficient Cooper Pair Splitting Without Interactions}
\author{E.S.~Tikhonov}
\email[e-mail:]{tikhonov@issp.ac.ru}
\affiliation{Osipyan Institute of Solid State Physics RAS, 142432 Chernogolovka, Russian Federation}
\affiliation{Condensed-Matter Physics Laboratory, HSE University, 101000, Moscow, Russia}
\author{V.S.~Khrapai}
\affiliation{Osipyan Institute of Solid State Physics RAS, 142432 Chernogolovka, Russian Federation}
\affiliation{National Research University Higher School of Economics, 20 Myasnitskaya Street, Moscow 101000, Russian Federation}
\begin{abstract}
For the three-terminal NSN device with single-mode normal terminals and without Coulomb blockade, we propose the interpretation of charge transfer process which allows us to consistently characterize the device operation as that of a Cooper pair splitter in terms of scattering matrix elements as well as in terms of measurable quantities. The obtained explicit expression for the splitting probability notably contains the two-particle interference term not available from conductance measurements. We show that splitting doesn't necessarily rely on single-particle crossed Andreev reflection amplitude thus allowing for the unit efficiency at zero energy. Our results imply that the current cross-correlator generally doesn't provide definite measure of splitting.
\end{abstract}
\maketitle

\section{Introduction}
Cooper pair splitter (CPS) is a solid-state device producing spin-entangled electron pairs. The original theoretical proposal~\cite{Recher2001} involved a superconductor coupled to two quantum dots in the Coulomb blockade regime, so that Cooper pairs constituting current through the superconductor are forced to split into the different leads of the setup. Following this idea, many theoretical and experimental works concentrate on the Coulomb blockaded devices~\cite{Borlin2002,Hofstetter2009,Herrmann2010,Golubev2010,Eldridge2010,Hofstetter2011,Burset2011,Schindele2012,Rech2012,Fulop2015,Tan2015,Borzenets2016,Dominguez2016,Baba2018,Walldorf2018,Walldorf2020,Ranni2021,Ranni2022,Scherubl2022,Bordoloi2022}. Hybrid NSN devices without Coulomb blockade are also thoroughly studied~\cite{Torres1999,Deutscher2000,Lesovik2001,Falci2001,Bignon2004,Beckmann2004,Russo2005,Kalenkov2007,
Melin2008,Golubev2009,Bednorz2009,Freyn2010,Wei2010,Das2012,Floser2013,Chen2015,Soori2017,Nehra2019,Golubev2019,Denisov2021,Denisov2022,Zsurka2023}, 
in particular with a goal to find out the manifestation of correlations, different microscopic processes and interactions in the measurable quantities. While the quantitative results obtained using various approaches coincide in certain limits, the interpretations sometimes differ significantly, down to whether Cooper pair splitting occurs or not~\cite{Golubev2010,Floser2013}.

The two most important operating characteristics of a CPS are (i) the splitting probability, $p_{11}$, determining the magnitude of the current of split pairs and (ii) the splitting efficiency, $K$, showing which part of the total superconductor current results in the separated entangled electrons. Unambiguous interpretation of the experiments requires the demonstration of the split electrons spin-entanglement as well as the extraction of~$K$. Usually, it is estimated from the analysis of the conductance correlations~\cite{Hofstetter2009,Herrmann2010,Hofstetter2011,Schindele2012,Fulop2015,Tan2015,Baba2018,Ranni2021,Ranni2022,Scherubl2022} with the recent experiments~\cite{Wang2022,Bordoloi2022,Bordin2023} performing also the spin correlation analysis. The existing proposals to achieve unit efficiency without Coulomb blockade rely on the energy filtering~\cite{Lesovik2001,Veldhorst2010,Burset2011,Sadovskyy2015} in order to maximize crossed Andreev reflection~(CAR). Both the possibility of reaching $K=1$ at zero energy and the corresponding role of CAR remain weakly explored.

Beyond the average current measurements, a valuable piece of information is contained in the current cross-correlations~\cite{Torres1999,Burkard2000,Melin2008,Freyn2010,Wei2010,Das2012,Floser2013,Golubev2019,Ostrove2019}. The basic logic suggests that for the dominating splitting events coincident arrival of electrons to two normal terminals would lead to the positive contribution to the cross-correlator. Since in the purely normal three-terminal Fermi-system cross-correlators are always negative, its positive sign in a superconductor-containing three-terminal device would then reflect the domination of the splitting events in the total current. However, the manifestation of splitting in the sign of the cross-correlator remains obscure since, e.g., for the case of transparent NSN junctions the well-established positive cross-correlator was shown to coexist with the suppressed~CAR~\cite{Floser2013}.

Here, we consider the three-terminal NSN device with \textit{single-mode} normal~(N) terminals in the zero-energy approximation. We show that the well-known general scattering matrix expressions for current correlators can be transformed in such a way as to reveal their connection with splitting processes. We obtain explicit expressions for splitting probability~$p_{11}$ and splitting efficiency~$K$ in terms of scattering matrix elements. The results contain two-particle interference term and thus are generally not accessible via the conductance measurements. We reveal that splitting processes do not rely solely on~CAR. We uncover the previously unexplored role of device geometry on the possibility of efficient splitting. Namely, the cases of \textit{single-mode} and \textit{multi-mode} superconducting~(S) terminals are conceptually different. For devices with single-mode superconducting~(S) terminal, we show that $K\leq1/2$ and that, counter-intuitively, the largest  $p_{11}$ and $K$ are realized at perfectly anti-correlated currents in two arms of the splitter. For devices with multi-mode S-terminal, we show that $K$~may reach unit without energy filtering. Finally, our results imply that the known value of the cross-correlator generally doesn't allow one to judge on~$p_{11}$ and~$K$.

\section{Approximations and main result}
\begin{figure}[h]
\begin{center}
\includegraphics[width=0.95\linewidth]{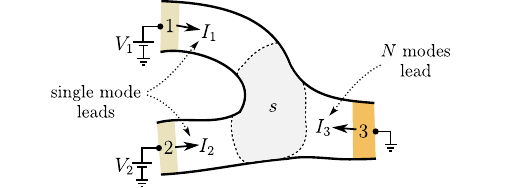}
\end{center}
\caption{Spin degenerate system with single-mode normal terminals~($1$ and $2$) and the superconducting~($3$) terminal. The central disordered region is described by the scattering matrix~$s$. The particular case of~$V_1=V_2$ effectively describes the splitter geometry.}
\label{fig1}
\end{figure}

In this manuscript we will study a spin-degenerate NSN~device with single-mode N-terminals depicted schematically in Fig.~\ref{fig1}. By single-mode terminals we mean that the arms connecting the terminals with the disordered region (shaded gray), described by the scattering matrix~$s$, are single-channel spin-degenerate conductors without scattering. The N-terminals are labeled by~$1$ and $2$, the S-terminal is labeled by~$3$. Throughout the paper, the upper greek letters denote the particle type, while the lower latin indices denote the terminals. In particular, $s_{11}^{eh}$ and $s_{22}^{eh}$ are the amplitudes of local Andreev reflection~(LAR), $s_{12}^{eh}$ and $s_{21}^{eh}$ describe CAR, $s_{11}^{ee}$, $s_{22}^{ee}$ and $s_{12}^{ee}$, $s_{21}^{ee}$ -- normal reflection and elastic cotunneling, respectively. The transmission coefficients, $T_{ij}^{\alpha\beta}=\left|s_{ij}^{\alpha\beta}\right|^2$, reflect the probabilities for a particle~$\beta$ from the terminal~$j$ to be transmitted as a particle~$\alpha$ to the terminal~$i$. Scattering matrix elements obey the orthogonality sum-rules~\cite{Anantram1996}
\begin{equation}
\sum\limits_{l\in N,S;\,\delta\in e,h}\left(s_{il}^{\alpha\delta}\right)^*s_{jl}^{\beta\delta}=\sum\limits_{l\delta}\left(s_{li}^{\delta\alpha}\right)^*s_{lj}^{\delta\beta}=\delta_{ij}\delta_{\alpha\beta},
\label{sumrules}
\end{equation}
with the following particular case for the transmission coefficients:
\begin{equation*}
\sum\limits_{j\in N,S;\,\delta\in e,h}T_{ij}^{\alpha\delta}=\sum\limits_{j\in N,S;\,\delta\in e,h}T_{ji}^{\delta\alpha}=1,
\end{equation*}
where the first sum includes all probabilities for a particle~$\alpha$ to find itself in an~$i$-th~terminal after the scattering event, while the second sum includes all scattering probabilities for a particle~$\alpha$ from the $i$-th~terminal. Currents flowing out of the terminals are taken to be positive. 

In our analysis we will rely on the general formulae for the average current and current correlations derived by Anantram and Datta for phase-coherent mesoscopic structures~\cite{Anantram1996}:
\small
\begin{equation*}
I_{i}=\frac{e}{h}\int dE \sum\limits_{\alpha,\,j\,\in\,NS,\,\beta}\text{sgn}(\alpha)\left[\delta_{ij}\delta_{\alpha\beta}-T_{ij}^{\alpha\beta}(E)\right]f_{j\beta}(E),
\end{equation*}
\begin{multline}
S_{ij}=\frac{2e^2}{h}\sum\limits_{\substack{k,\,l\,\in\, N,S\\ \alpha,\,\beta,\,\gamma,\,\delta\,\in\, e,\,h}}\text{sgn}(\alpha)\text{sgn}(\beta)\times \\ \times\int dE\,A_{k\gamma;\,l\delta}(i\alpha,E)A_{l\delta;\,k\gamma}(j\beta,E)f_{k\gamma}(E)\left[1-f_{l\delta}(E)\right],
\label{gennoise}
\end{multline}
\normalsize
where $A_{k\gamma;\,l\delta}(i\alpha,E)=\delta_{ik}\delta_{il}\delta_{\alpha\gamma}\delta_{\alpha\delta}-\left(s_{ik}^{\alpha\gamma}\right)^{*}s_{il}^{\alpha\delta}$, $f_{i\alpha}$ are the Fermi distribution functions for electrons and holes in the reservoirs limiting in our case of zero temperature to Heaviside step functions. 

In the following we operate in the sub-gap energy range assuming no quasiparticle loss in the superconductor,
which leaves only N-terminals indices in the $s$-matrix and, in particular, allows us to omit summation over the S-terminal index in sum rules. Additionally, we assume one may neglect the energy dependence of the scattering matrix in the considered energy range.
The expressions for average currents linearize to (see equation~(31) in~\cite{Anantram1996})
\begin{align*} 
I_1 &= G_0\left[V_1\left(1-T_{11}^{ee}+T_{11}^{he}\right)+V_2\left(-T_{12}^{ee}+T_{12}^{he}\right)\right], \\ 
I_2 &= G_0\left[V_1\left(-T_{21}^{ee}+T_{21}^{he}\right)+V_2\left(1-T_{22}^{ee}+T_{22}^{he}\right)\right].
\end{align*}
Consider the special case of $V_1=V_2=V$, effectively describing the splitter geometry. Using the sum rule and taking into account the electron-hole symmetry, for the dimensionless average currents, ${\cal I}=I/(G_0V)$ with $G_0=2e^2/h$, one gets
\begin{align}
\begin{split}
{\cal I}_1 &=1-T_{11}^{ee}+T_{11}^{he}-T_{12}^{ee}+T_{12}^{he}=2\left(T_{11}^{eh}+T_{12}^{eh}\right), \\
{\cal I}_2 &=1-T_{22}^{ee}+T_{22}^{he}-T_{21}^{ee}+T_{21}^{he}=2\left(T_{21}^{eh}+T_{22}^{eh}\right).
\end{split}
\label{avcurs}
\end{align}

Generally utilized way to proceed with expressions for current correlators from~(\ref{gennoise}) is to explicitly write down all its numerous components~\cite{Torres1999,Torres2001,Floser2013}. These components are commonly identified with specific microscopic processes, e.g. the term containing the product~$s_{12}^{ee}s_{21}^{ee}(s_{11}^{ee})^{\dagger}(s_{22}^{ee})^{\dagger}$ is put in correspondence with the combination of elastic cotunneling and normal reflection, while the term containing the product~$s_{21}^{he}s_{12}^{he}(s_{11}^{he})^{\dagger}(s_{22}^{he})^{\dagger}$ is attributed to the combination of LAR and CAR. Further conclusion on the contribution from splitting to the cross-correlator~$S_{12}$ is based on the input from the components containing CAR amplitudes~\cite{Freyn2010,Floser2013}. Here, we argue that in fact CAR and splitting are generally not directly related and, moreover, splitting does not necessarily lead to the positive contribution to~$S_{12}$. Our claim is based on the zero-energy transformation of the expressions~(\ref{gennoise}) for current correlators which demonstrates the two-particle interference nature of splitting. Note that in our claim we do not oppose the specific quantitative results obtained previously by other authors some of which will be reproduced below. Rather, we show that the interference form of the transformed expressions allows the novel elegant interpretation which may be not as straightforward as was previously believed.

Similarly to the average currents, further we will consider the dimensionless zero-frequency current correlators, ${\cal S}_{ij}=S_{ij}/(G_0eV)$. In Supplemental Material~(SM) sections~I-III we analytically demonstrate that for the system under consideration the expressions~(\ref{gennoise}) can be rewritten as follows:
\begin{equation}
\begin{cases}
{\cal S}_{11}=2{\cal I}_1(2-{\cal I}_1)-4\left|\twoprodpr{1}{1}{e}{e}{1}{2}{h}{e}-\twoprodpr{1}{1}{h}{e}{1}{2}{e}{e}\right|^2, \\[8pt]
{\cal S}_{22}=2{\cal I}_2(2-{\cal I}_2)-4\left|\twoprodpr{1}{1}{e}{e}{1}{2}{h}{e}-\twoprodpr{1}{1}{h}{e}{1}{2}{e}{e}\right|^2, \\[8pt]
{\cal S}_{12}=4\left|\twoprodpr{1}{1}{e}{e}{1}{2}{h}{e}-\twoprodpr{1}{1}{h}{e}{1}{2}{e}{e}\right|^2-2{\cal I}_1{\cal I}_2+\\[8pt]
\quad\quad+8\left|s_{12}^{he}s_{21}^{he}-s_{11}^{he}s_{22}^{he}\right|^2.
\end{cases}
\label{expr1}
\end{equation}
Note the occurrence of two-particle interference terms not reduced to average currents. The common to all three expressions term remains the same with the interchange of terminals indices, 
\begin{multline*}
\twoprodpr{1}{1}{e}{e}{1}{2}{h}{e}-\twoprodpr{1}{1}{h}{e}{1}{2}{e}{e}=\twoprod{1}{1}{h}{h}{1}{2}{h}{e}+\twoprod{1}{1}{e}{h}{1}{2}{e}{e}=\\
=-\left\{\twoprod{2}{1}{e}{h}{2}{2}{e}{e}+\twoprod{2}{1}{h}{h}{2}{2}{h}{e}\right\}=\twoprodpr{2}{2}{e}{e}{2}{1}{h}{e}-\twoprodpr{2}{2}{h}{e}{2}{1}{e}{e},
\end{multline*}
which is the consequence of the orthogonality rule~(\ref{sumrules}). Below we will show that~(\ref{expr1}) allows for a transparent interpretation of how current flows in a device. It also allows for the introduction of explicit expressions for splitting probability and efficiency which are our main result:
\begin{equation}
p_{11}=2\left|\twoprodpr{1}{1}{e}{e}{1}{2}{h}{e}-\twoprodpr{1}{1}{h}{e}{1}{2}{e}{e}\right|^2,\quad K=2p_{11}/({\cal I}_1+{\cal I}_2)
\end{equation} 
In SM section~IV we prove that generally
\begin{equation}
p_{11}\leq1/2 \quad \text{and}\quad K\leq 1-|{\cal I}_3|/4,
\label{mybounds}
\end{equation}
where ${\cal I}_3$ is the total superconductor current. In the following we will first address the case of single-mode S-terminal and then move on to the multi-mode S-terminal problem.

\section{Single-mode S-terminal}
We start our analysis with the device with single-mode S-terminal. This implies the following crucial identity for the Andreev amplitudes:
\begin{equation}
s_{12}^{he}s_{21}^{he}=s_{11}^{he}s_{22}^{he},
\label{identity}
\end{equation}
which can be derived by connecting the scattering matrix to that of a complementary device with all N-terminals, see SM~section~V. The corresponding expression for cross-correlators are therefore
\begin{equation}
\begin{cases}
{\cal S}_{11}=2{\cal I}_1(2-{\cal I}_1)-4\left|\twoprodpr{1}{1}{e}{e}{1}{2}{h}{e}-\twoprodpr{1}{1}{h}{e}{1}{2}{e}{e}\right|^2, \\[8pt]
{\cal S}_{22}=2{\cal I}_2(2-{\cal I}_2)-4\left|\twoprodpr{1}{1}{e}{e}{1}{2}{h}{e}-\twoprodpr{1}{1}{h}{e}{1}{2}{e}{e}\right|^2, \\[8pt]
{\cal S}_{12}=4\left|\twoprodpr{1}{1}{e}{e}{1}{2}{h}{e}-\twoprodpr{1}{1}{h}{e}{1}{2}{e}{e}\right|^2-2{\cal I}_1{\cal I}_2.
\end{cases}
\label{singlemodenoise}
\end{equation}

\begin{figure}[h]
\begin{center}
\includegraphics[width=\linewidth]{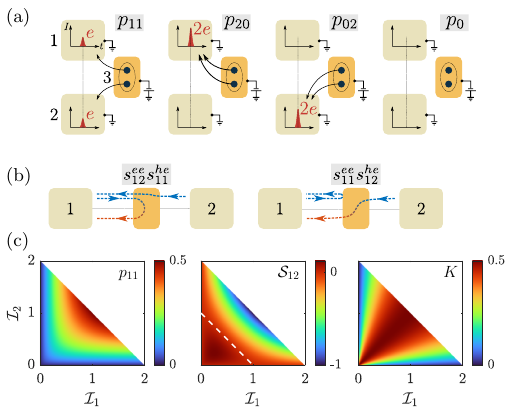}
\end{center}
\caption{(a)~Possible outcomes of the current pulse in a single-mode NSN~device. (b)~Two interfering processes contributing to~$p_{11}$. Blue and red color correspond to electron and hole states, respectively. (c)~Splitting probability, current cross-correlator and the efficiency of splitting in a single-mode NSN-device as functions of~${\cal I}_1$ and ${\cal I}_2$. The dashed line on the central panel corresponds to~${\cal S}_{12}=0$. Only colored points are available and any of these points is achievable in, e.g., the quantum Hall Bogoliubov interferometer (see text).}
\label{fig2}
\end{figure}

Consider now a physical picture of the current flowing process in the system under consideration. For a quantum conductor at low enough temperature it is a consequence of the Pauli principle that, up to logarithmically small fluctuations, each terminal periodically, with a frequency of $eV/h$ per spin, attempts to emit a particle into each mode of a conductor~\cite{LL1993}. In our case, from the point of view of the superconductor, there are four possible current pulse outcomes, schematically depicted in Fig.~\ref{fig2}(a). Eventually, the Cooper pair is either split into two different arms connected to normal terminals (with the splitting probability~$p_{11}$), or both charges are transmitted into the same arm ($p_{20}$ and $p_{02}$), or the attempt is unsuccessful and no charge is transmitted ($p_0$). While the splitting probability~$p_{11}$ defines the magnitude of the current of split pairs, the splitting efficiency~$K$ shows the fraction of the total superconductor current which is split.

The introduced probabilities obey the normalization condition and determine the currents:
\begin{equation*}
\begin{cases}
1=p_0+p_{11}+p_{02}+p_{20}, \\[8pt]
{\cal I}_1=p_{11}+2p_{20}, \\[8pt]
{\cal I}_2=p_{11}+2p_{02}.
\end{cases}
\end{equation*}
Moreover, for the current correlators in the terminals, one obtains 
\begin{equation*}
\begin{cases}
{\cal S}_{11}=2[p_{11}+4p_{20}-\left(p_{11}+2p_{20}\right)^2], \\[8pt]
{\cal S}_{22}=2[p_{11}+4p_{02}-\left(p_{11}+2p_{02}\right)^2], \\[8pt]
{\cal S}_{12}=2[p_{11}-\left(p_{11}+2p_{20}\right)\left(p_{11}+2p_{02}\right)].
\end{cases}
\end{equation*}
Here, we used the following expression for the zero-frequency current correlators: $S_{ij}=2\langle\delta Q_i\delta Q_j\rangle/t$, where $t$ -- is the long enough observation time and $\delta Q_i$ is the fluctuation of charge~$Q_i$ which passes the cross-section of the lead connected to the~$i$-th terminal during this time. To bridge these relations with the scattering matrix approach, we rewrite them in a slightly different form leaving only the splitting probability:
\begin{equation}
\begin{cases}
{\cal S}_{11}=2{\cal I}_1(2-{\cal I}_1)-2p_{11}, \\[8pt] {\cal S}_{22}=2{\cal I}_2(2-{\cal I}_2)-2p_{11}, \\[8pt] {\cal S}_{12}=2p_{11}-2{\cal I}_1{\cal I}_2.
\end{cases}
\label{expr2}
\end{equation}

We first test these relations for the well-known two-terminal case of the NS-junction where the lead to the second terminal is off. One therefore has to take $p_{11}=p_{02}=0$, so that the device is effectively described by any of the two other probabilities with $p_{0}+p_{20}=1$ and ${\cal I}=2p_{20}$. Here, $p_{20}$ is the LAR~probability expressed via the normal-state transmission probability~$T$ via the standard relation $p_{20}=T^2/(2-T)^2$. For the conductance and the noise auto-correlator one then obtains $G_{\text{NS}}=(4e^2/h)p_{20}$ and ${\cal S}_{\text{NS}}={\cal S}_{11}=4p_0{\cal I}$. In a particular case of a transparent NS-boundary, $p_0=0$, the conductance is $G_{\text{NS}}=4e^2/h$ and the current flow is noiseless which verifies our assumption of periodical current pulse attempts for a superconductor. In the opposite limit of a tunnel NS-junction, $p_{0}\to1$, the auto-correlator is ${\cal S}=4{\cal I}$, reproducing the doubling of shot noise compared to the Poisson value ${\cal S}_{\text{P}}=2{\cal I}$, characteristic for the normal state tunnel junction~\cite{DeJong1994,Kozhevnikov2000,Jehl2000}.

Returning to the case of a three-terminal device, we note that~(\ref{expr2})~provides~$p_{11}$, and hence~$K$, in terms of the measurable quantities. Comparing (\ref{singlemodenoise}) and (\ref{expr2}) we obtain the following explicit expressions for the splitting probability:
\begin{equation*}
p_{11}=2\left|\twoprodpr{1}{1}{e}{e}{1}{2}{h}{e}-\twoprodpr{1}{1}{h}{e}{1}{2}{e}{e}\right|^2=2\left|\twoprodpr{2}{2}{e}{e}{2}{1}{h}{e}-\twoprodpr{2}{2}{h}{e}{2}{1}{e}{e}\right|^2.
\end{equation*}
These expressions suggest the splitting events include the two-particle interference between two processes with indistinguishable results shown in Fig.~\ref{fig2}(b). Note that only for the convenience we illustrate these processes in a geometry where S-terminal supports at least two modes.

The relation for the Andreev amplitudes~(\ref{identity}) allows one to transform~$p_{11}$ as follows, see SM~section~VI:
\begin{equation*}
p_{11}=2T_{21}^{he}\frac{T_{12}^{he}+T_{11}^{he}}{T_{21}^{he}+T_{11}^{he}}=2T_{12}^{he}\frac{T_{21}^{he}+T_{22}^{he}}{T_{12}^{he}+T_{22}^{he}}.
\end{equation*}
In particular, in the absence of a magnetic field, $p_{11}$ reduces to the CAR probability, $p_{11}=2T_{12}^{he}=2T_{21}^{he}$. Generally, under broken time-reversal symmetry, $p_{11}$ depends also on the LAR~probabilities. For the symmetric device, ${\cal I}_1={\cal I}_2$ ($p_{20}=p_{02}$), where (\ref{identity})~implies $T_{12}^{he}=T_{22}^{he}$, $T_{11}^{he}=T_{21}^{he} $, the splitting probability may be written as $p_{11}=T_{12}^{he}+T_{21}^{he}$. 

\begin{figure}[h]
\begin{center}
\includegraphics[width=\linewidth]{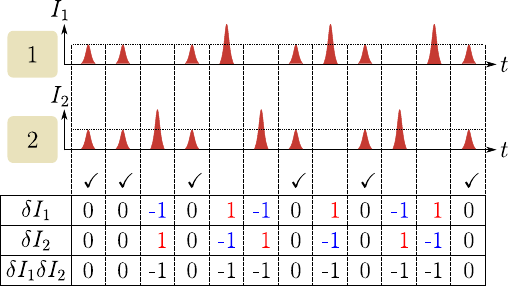}
\end{center}
\caption{Time dependence of current in two arms of a device with single-mode S-terminal where splitting processes (marked with checks) provide maximum possible split current. Splitting processes do not contribute to the current cross-correlator. Maximum possible split current coexists with perfectly anti-correlated currents.
}
\label{fig_illust_neg}
\end{figure}

We notice that for the discussed device $K$, all the probabilities and the current correlators may be expressed in terms of the currents. In particular:
\begin{equation*}
K=\frac{2{\cal I}_1{\cal I}_2}{({\cal I}_1+{\cal I}_2)^2},\quad p_{11}=\frac{{\cal I}_1{\cal I}_2}{{\cal I}_1+{\cal I}_2}, 
\end{equation*}
\begin{equation*}
{\cal S}_{12}=2{\cal I}_1{\cal I}_2/({\cal I}_1+{\cal I}_2)-2{\cal I}_1{\cal I}_2,
\end{equation*}
see Fig.~\ref{fig2}(c) summarizing the values of~$p_{11}$, ${\cal S}_{12}$ and $K$ as functions of the normalized currents in N-terminals. The above expressions resolve the whole class of splitters with single-mode S-terminal. Importantly, the sign of the cross correlator is determined simply by the total current. Moreover, one has the following bounds:
\begin{equation*}
K\leq\frac12,\quad p_{11}\leq\frac12,
\end{equation*}
that is only half of the superconductor current can in principle be split which is the consequence of~(\ref{identity}).   Maximization of the split current requires ${\cal I}_1={\cal I}_2=1$, which in terms of probabilities reduces to $p_{11}=1/2$, $p_{20}=p_{02}=1/4$, $p_{0}=0$, coming along with the following N-terminals noise correlators: ${\cal S}_{11}={\cal S}_{22}=1$, ${\cal S}_{12}=-1$. Therefore the largest possible~$p_{11}$ and $K$ coexist with the largest possible negative~${\cal S}_{12}$. In fact, the currents in a so-tuned setup are anti-correlated:
\begin{equation*}
\langle\left(\delta {\cal I}_1+\delta {\cal I}_2\right)^2\rangle={\cal S}_{11}+{\cal S}_{22}+2{\cal S}_{12}=0\to \delta{\cal I}_1=-\delta{\cal I}_2.
\end{equation*}
This observation contradicts the general basic logic which would require maximizing cross-correlator to achieve maximum splitting current and/or efficiency. The physical reason for this seemingly counter-intuitive result is that for the above probabilities the momentary currents in a splitting event equal the average currents and do not contribute to~${\cal S}_{12}$, see Fig.~\ref{fig_illust_neg}.  

\subsection{Beam Splitter}
\begin{figure}[h]
\begin{center}
\includegraphics[width=\linewidth]{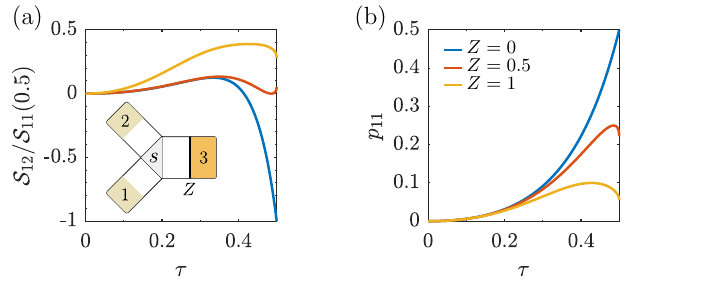}
\end{center}
\caption{(a)~Normalized current cross-correlator and (b)~splitting probability in a beam splitter geometry for different values of disorder at S-interface.
}
\label{fig_BS}
\end{figure}

We now consider the cases where $s$-matrix can be calculated analytically. As the simplest application of our findings, consider first the geometry of beam splitter (BS)~\cite{Torres1999} shown in the inset of Fig.~\ref{fig_BS}. Here, the scattering region~(gray) is represented by the semitransparent mirror so that the device is symmetric and is parametrized by the single parameter~$0<\tau<1/2$ which controls the transparency of the splitter: the transmission between the S and the N terminals vanishes at~$\tau=0$ and is maximal at~$\tau=1/2$. Additionally, the interface between S-terminal and the corresponding lead (black thick line) is disordered and is characterized by the BTK~parameter~$Z$ related to the interface transparency via $T=1/(1+Z^2)$~\cite{Blonder1982}. The scattering matrix of the device is readily available~\cite{Torres2001} and provides one with explicit analytical expressions for currents~${\cal I}_1$, ${\cal I}_2$, current correlators~${\cal S}_{11}$, ${\cal S}_{12}$ and splitting probability~$p_{11}$ and efficiency~$K$, see SM~section~VII. Importantly, the symmetry of the device at any~$Z$ ensures that the efficiency is exactly~$50\%$. For the case of transparent interface, $Z=0$, the analytical expressions take simple form:
\begin{equation*}
{\cal I}_1={\cal I}_2=\tau^2/(1-\tau)^2, \,\, {\cal S}_{12}=\tau^2(1-2\tau-\tau^2)/(1-\tau)^4,
\end{equation*}
\begin{equation*}
p_{11}=\tau^2/\left[2(1-\tau)^2\right].
\end{equation*}
Note that up to a factor of~$2$ coming from the definition, the expression for~${\cal S}_{12}$ is exactly the same as in~\cite{Torres1999}, however now we have revealed that it is determined just by the total current of the superconductor:
\begin{equation*}
{\cal S}_{12}={\cal I}_1-2{\cal I}_1^2={\cal I}_3\left(1-{\cal I}_3\right)/2.
\end{equation*}

In Fig.~\ref{fig_BS} we demonstrate the dependences of normalized~${\cal S}_{12}$ and $p_{11}$ on mirror transparency for~$Z=0$, $0.5$ and $1$. Note that the curve for~${\cal S}_{12}|_{Z=1}$ is almost the same as that by J. Torres and T. Martin plotted for~$V/\Delta=0.5$, see Fig.~3 (top curve)~\cite{Torres1999} with the difference due to the finite bias used in~\cite{Torres1999}. As is dictated by our general analysis, the maximum possible split current achieved at $Z=0$, $\tau=1/2$ coexists with perfectly anti-correlated currents. At the same time, the efficiency of splitting is both $Z$- and $\tau$-independent, $K=1/2$, since the device is symmetric. Note also that at high enough disorder at the S-interface the total superconductor current is ${\cal I}_3\ll1$, so that ${\cal S}_{12}\approx2p_{11}={\cal I}_3/2$. In this limit the correspondence between splitting and the cross-correlator is valid since current and its fluctuations are almost identical.

\subsection{Quantum Hall Bogoliubov interferometer}
\begin{figure}[h]
\begin{center}
\includegraphics[width=\linewidth]{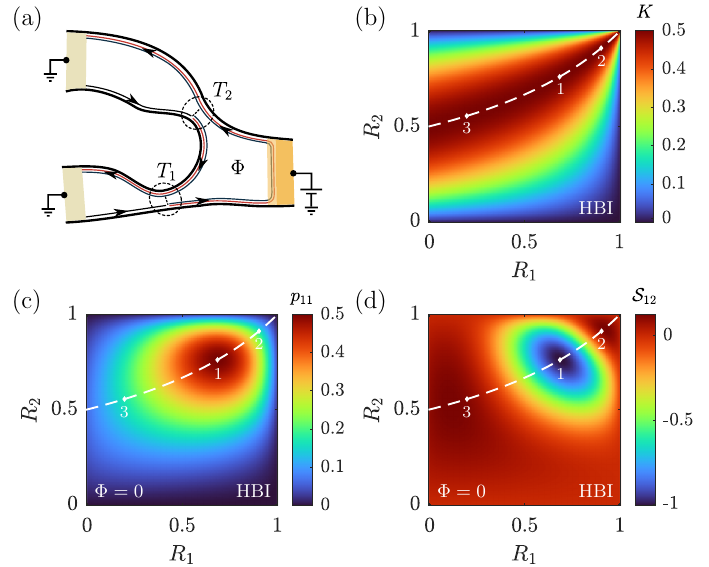}
\end{center}
\caption{(a)~Sketch of the quantum Hall Bogoliubov interferometer. (b)~$\Phi$-independent splitting efficiency of the HBI. (c,d)~Splitting probability and current cross-correlator for the~HBI at $\Phi=0$, $\left|t_{ee}\right|^2=0.9$.
}
\label{fig_HBI}
\end{figure}

We now discuss the more complicated case of the recently introduced quantum Hall Bogoliubov interferometer (HBI)~\cite{Khrapai2023}. The system is a modification of a Fabry-Perot quantum Hall interferometer~\cite{DeC.Chamon1997} with a superconducting terminal inside. For the HBI at the filling factor~$\nu=2$, the scattering matrix can be obtained analytically, see SM~section~VIII. In particular, maximizing the $\Phi$-independent splitting efficiency, $K=2T_1T_2R_2/(1-R_1R_2)^2=1/2$, requires $T_2=T_1/(1+T_1)$, see the dashed line in Fig.~\ref{fig_HBI}(b). Here, $T_1$ and $R_1$ ($T_2$ and $R_2$) are transmission and reflection probabilities of the two constrictions, $\Phi$ is the Aharonov-Bohm phase. We note that the obtained condition simultaneously ensures the symmetric device. For any fixed constrictions, the values of $p_{11}$, ${\cal I}_1$ and ${\cal I}_2$ achieve their largest values at $\Phi=2\pi m$, $m\in \mathbb{Z}$. Figs.~\ref{fig_HBI}(c) and \ref{fig_HBI}(d) demonstrate $p_{11}$ and ${\cal S}_{12}$ for this specific case at $|t_{ee}|^2=0.9$ ($t_{ee}$ and $t_{eh}$ are the amplitudes of normal and Andreev scattering off the S-terminal). The dashed lines here are the same as in panel~(b). In particular, at $T_1=\left|t_{eh}\right|$ and $T_2=\left|t_{eh}\right|/(1+\left|t_{eh}\right|)$, which is indicated by point~$1$ in panels (b)-(d), both $K$ and $p_{11}$ are maximal: $K=1/2$, $p_{11}=1/2$. At the same time, ${\cal I}_1={\cal I}_2=1$ and ${\cal S}_{11}={\cal S}_{22}=1$, ${\cal S}_{12}=-1$, so that the currents here are anti-correlated. Note that by tuning~$\Phi$ one can vary the current of split pairs saving the value of~$K$, see SM~Fig.2.

\section{Multi-mode S-terminal} 
We emphasize that the whole above discussion strongly relies on the fact that exactly one mode connects the S-terminal to the scattering region. To go beyond this geometry, we assume there are at least two modes between the S-terminal and the scattering region. The expressions~(\ref{expr1}) do not change since the scattering matrix~$s$ includes only N-terminals indices. However, the relation~(\ref{identity}) between AR amplitudes is no longer valid: the simplest example is when two N-terminals are connected to the large superconductor with perfectly transparent interfaces in two significantly spatially separated places with no disorder so that $s_{12}^{he}=s_{21}^{he}=0$, while $|s_{11}^{he}|=|s_{22}^{he}|=1$. As a result, compared to the case of single-mode S-terminal the current cross-correlator gets an additional modulus-squared term while the expressions for auto-correlators remain the same. Denoting 
\begin{equation*}
p_{22}=|s_{12}^{he}s_{21}^{he}-s_{11}^{he}s_{22}^{he}|^2,
\end{equation*}
we write
\begin{equation}
\begin{cases}
{\cal S}_{11}=2{\cal I}_1(2-{\cal I}_1)-2p_{11}, \\[8pt] {\cal S}_{22}=2{\cal I}_2(2-{\cal I}_2)-2p_{11}, \\[8pt] {\cal S}_{12}=2p_{11}-2{\cal I}_1{\cal I}_2+8p_{22}.
\end{cases}
\label{expr3}
\end{equation}
Note that for the case of single-mode S-terminal geometry, destructive interference leads to~$p_{22}=0$.

\begin{figure}[h]
\begin{center}
\includegraphics[width=\linewidth]{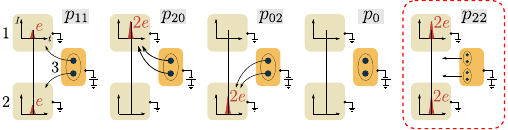}
\end{center}
\caption{Possible outcomes of a current pulse in a device with multi-mode S-terminal.}
\label{fig_p22}
\end{figure}

\begin{figure}[h]
\begin{center}
\includegraphics[width=\linewidth]{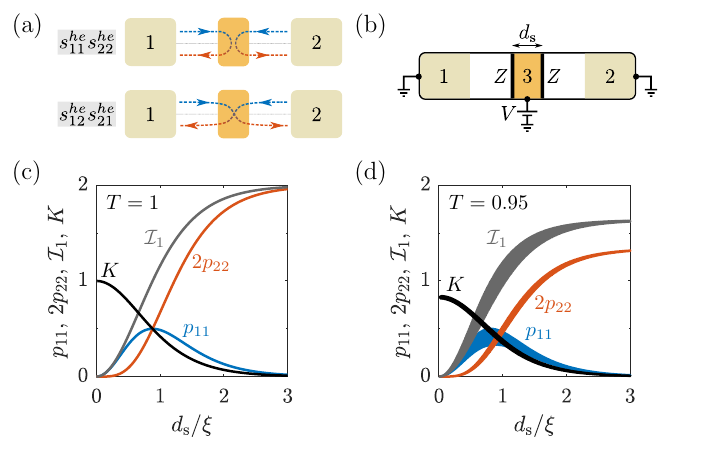}
\end{center}
\caption{(a)~Two interfering processes contributing to~$p_{22}$. (b)~NSN hybrid device with S-region connected to two single-mode N-terminals with interfaces characterized by BTK parameter~$Z$. (c)~Probabilities~$p_{11}$ and $2p_{22}$, current in one of the arms~${\cal I}_1$ and efficiency~$K$ for NSN device from panel~(b) as functions of the normalized S-region length for S-interfaces transparency~$T=1$. (d)~Same as in panel~(c) for the case of~$T=0.95$. We used $k_{\text{F}}=1.36\cdot10^{10}\,\text{m}^{-1}$ for the calculations.
}
\label{fig4}
\end{figure}

The above expressions describe the random process with the one new possible event in addition to those depicted in Fig.~\ref{fig2}(a). This new event results in the simultaneous arrivals of two electrons to both N-terminals and is described by the probability~$p_{22}$, see Fig.~\ref{fig_p22} for the schematic representation of this event. Similar to the splitting events, $p_{22}$ includes the two-particle interference between two processes with indistinguishable results, see Fig.~{\ref{fig4}}(a). We emphasize significant difference between contributing to~$p_{11}$ processes $s_{12}^{ee}s_{11}^{he}$ and $s_{11}^{ee}s_{12}^{he}$, as well as between contributing to~$p_{22}$ processes $s_{11}^{he}s_{22}^{he}$ and $s_{12}^{he}s_{21}^{he}$: in the absence of disorder the former ones are possible while the latter ones are not.

\subsection{Line-shaped NSN-device}
Consider as an example the geometry of a line-shaped NSN-device with single-mode N-terminals and the S-region of length~$d_{\text{s}}$ with the coherence length~$\xi$, see~Fig.~\ref{fig4}(b). Both S-interfaces are assumed identical and are characterized by the BTK parameter~$Z$. This geometry was thoroughly studied previously~\cite{Melin2008,Golubev2010,Freyn2010,Floser2013}. Generally, for the case of not perfectly transparent S-interfaces with~$Z>0$, propagation through the superconductor in this one dimensional model includes rapidly oscillating factors with the period of~$\lambda_{\text{F}}\ll d_{\text{s}}$~\cite{Falci2001}. In~SM~Fig.~3 we demonstrate oscillations of ${\cal S}_{12}$ and $p_{11}$ with~$d_{\text{s}}$. In SM~Figs.~(4,5) we verify that after averaging our expressions for current correlators~(\ref{expr3}) perfectly reproduce analogous results obtained by other authors~\cite{Freyn2010,Floser2013}. In the following we concentrate on the most interesting limit of both S-interfaces near to perfectly transparent.

We start with the case of ideal interfaces, $T=1$ ($Z=0$), where the aforementioned oscillating factors do not come into play. The scattering matrix reads~\cite{Entin-Wohlman2008}

\begin{center}
$s_{11}^{ee}=0$, $s_{12}^{ee}=1/\cosh(d_{\text{s}}/\xi)$,

$s_{12}^{he}=s_{21}^{he}=0$, $s_{11}^{he}=s_{22}^{he}=i\tanh(d_{\text{s}}/\xi)$,
\end{center}

providing $p_{20}=p_{02}=0$, $p_0=1/\cosh^4(d_{\text{s}}/\xi)$ and  
\begin{equation*}
p_{11}=\frac{2\sinh^2(d_{\text{s}}/\xi)}{\cosh^4(d_{\text{s}}/\xi)},\quad p_{22}=\tanh^4(d_{\text{s}}/\xi).
\end{equation*}
\begin{equation*}
{\cal I}_1={\cal I}_2=2\tanh^2(d_{\text{s}}/\xi),
\end{equation*}
\begin{equation*}
{\cal S}_{11}={\cal S}_{12}=\frac{4\sinh^2(d_{\text{s}})/\xi}{\cosh^4(d_{\text{s}})/\xi}.
\end{equation*}
Note the limiting case of our expressions, ${\cal S}_{11}={\cal S}_{12}=4(d_{\text{s}}/\xi)^2$ at $d_{\text{s}}\ll\xi$, coinciding with the result for the perfect interfaces from~\cite{Freyn2010}, see Appendix~B. The dependences on~$d_{\text{s}}/\xi$ of~${\cal I}_1$ along with contributing to it~$p_{11}$ and $2p_{22}$ is demonstrated in~Fig.~\ref{fig4}(c). Note that while CAR is absent, $s_{12}^{he}=s_{21}^{he}=0$, splitting events occur as a result of two-particle interference, see the left part of Fig.~\ref{fig2}(b). Importantly, at small enough $d_{\text{s}}\lesssim\xi$, it is splitting events that dominate in the total current of the superconductor, ${\cal I}_3=2(p_{11}+2p_{22})$. The maximum~$p_{11}=1/2$ is achieved around~$d_{\text{s}}\approx\xi$ and the further increase of the superconductor length predictably leads to the decrease of $p_{11}$ (and $K$) and the domination of $2p_{22}$. This observation reveals LAR events at two S-interfaces. Fig.~\ref{fig4}(c) also shows the efficiency, $K=1/\cosh^2(d_{\text{s}}/\xi)$, which in the limit of $d_{\text{s}}\ll\xi$ goes to unity, importantly, at zero energy and without energy filtering~\cite{Sadovskyy2015}. The observation of $K=1$ at $d_{\text{s}}/\xi\to0$ is reasonable since in this case all transmission probabilities equal zero besides $T_{11}^{he}=T_{22}^{he}=T\ll1$ and $T_{12}^{ee}=T_{21}^{ee}=1-T\approx1$. The processes depicted in the left side of Fig.~\ref{fig2}(b) are therefore the only ones contributing to the current flow (along with the processes corresponding to $s_{21}^{ee}s_{22}^{he}$). Interestingly, for the discussed device as well as for any setup without CAR ($s_{12}^{he}=s_{21}^{he}=0$), the cross-correlator is ${\cal S}_{12}=2p_{11}>0$ and provides exactly information about the splitting processes. However, in the general case of present CAR there is no obvious connection between splitting processes and the value of~${\cal S}_{12}$. Finally, we mention that the case of ideal interfaces delivers an upper boundary on efficiency from~(\ref{mybounds}). 

Consider now the case of the finite transparency of both S-interfaces. Here, $Z>0$ leads to the rise of oscillations in the scattering matrix elements resulting also in the oscillations in average currents, current correlators and probabilities. It is these oscillations that lead to the previously discussed negative values of ${\cal S}_{12}$ at intermediate interfaces transparencies~\cite{Freyn2010,Floser2013} as we demonstrate in SM~Fig.~3. Here in the main text, we concentrate on the case of nearly transparent interfaces where these oscillations are not too strong. The analytical expressions are obtained symbolically using Appendix~A from~\cite{Freyn2010} to first derive the scattering matrix elements which are then used to compute average currents and current correlators with~(\ref{avcurs}) and~(\ref{expr3}), respectively. We demonstrate our results in Fig.~\ref{fig4}(d) for the case of~$T=0.95$. At a given value of~$d_{\text{s}}/\xi$, the current in each arm of the device decreases compared to the case of perfect interfaces. At the same time, $p_{20}$ and $p_{02}$ come into play reducing the efficiency of splitting. The broadened lines represent the oscillations in ${\cal I}_1$ and ${\cal S}_{12}$, as well as in $p_{11}$ and $p_{22}$, which are due to the finite transparency of interfaces (the corresponding ${\cal S}_{12}$ is demonstrated in SM~Fig.~6). At large enough $d_{\text{s}}/\xi$, the oscillations in ${\cal I}_1$ die out and it correctly limits to the value dictated by the probability of LAR,
\begin{equation*}
{\cal I}_1\approx2p_{\text{LAR}}=2T^2/(2-T)^2\approx1.64.
\end{equation*}
For such long devices splitting is negligible and we observe that $p_{22}\approx p_{\text{LAR}}^2$, indicating that it describes simultaneous statistically independent LAR events at two S-interfaces.

\section{Discussion and conclusion}
Several remarks stem from the discussion of the line-shaped NSN~device and are due to be made. Our results for average currents and current correlators exactly coincide with the results obtained previously also using the scattering matrix formalism~\cite{Freyn2010,Floser2013}, compare our SM~Figs.~(3,4) with Fig.~3 from~\cite{Freyn2010}. Importantly, in the limit under consideration, the result obtained using Keldysh Green functions technique~\cite{Golubev2010} is also consistent with the result of~\cite{Freyn2010} and therefore with our result, see the discussion right after equation~(37) in~\cite{Golubev2010}. While the results coincide, the three interpretations are different. The authors of~\cite{Golubev2010} conclude that the positive cross-correlations are due to CAR.
On the contrary, the authors of~\cite{Floser2013} reason that since at $Z=0$ CAR processes do not dominate neither in the conductance nor in the noise, the positive sign of~${\cal S}_{12}$ should not be ascribed to Cooper pair splitting. Important to note here is the implicit identification of CAR and Cooper pair splitting in both cases. Our interpretation is different from both of these. Namely, it is the two-particle processes depicted in~Fig.~\ref{fig2}(b) that result in splitting rather than single particle CAR~($s_{12}^{he}$ or $s_{21}^{he}$). Our analysis suggests that the splitting probability~$p_{11}$ is not determined solely by CAR amplitude but rather is determined by the interference of these two processes. Note also that in the absence of time-reversal symmetry different CAR amplitudes are not necessarily the same as, e.g., in the case of HBI, see SM~section~VIII. In particular, at~$R_1=R_2=0$ and $|t_{eh}|=1$ we obtain $|s_{21}^{he}|=1$, $s_{12}^{he}=0$ and no splitting. This observation in itself demonstrates that identification of splitting with CAR is generally meaningless.

In summary, we describe operation of the NSN-based CPS with single-mode N-terminals and demonstrate the possibility of unit splitting efficiency without any energy filtering. The provided examples underline the necessary steps to make further progress. First, the generalization of our explicit $s$-matrix-based expressions for the case of multi-mode N-terminals may shed light on whether efficient Cooper pair splitting is possible in realistic devices with multi-mode N-terminals. Note that on the technological side this type of devices will also require almost transparent S-interfaces. HBI-like devices, at filling factor $\nu>2$, may provide an appealing alternative since the number of channels is controllable by the magnetic field, Coulomb effects can me minimized by the metallic back gate due to the shunting capacitance~\cite{Golubev2010}, and the S-interface quality is not as crucial as for the line-shaped devices.

We acknowledge valuable discussions with D.V.~Shovkun. The work was supported by the Russian Science Foundation project 22-12-00342. The analysis of line-shaped device with imperfect interfaces [Fig.~7(d)] was supported by the Basic research program of HSE.

\clearpage
\begin{widetext}
\begin{center}
\textbf{\large Supplemental Material}
\end{center}
\setcounter{equation}{0}
\setcounter{figure}{0}
\setcounter{table}{0}
\setcounter{section}{0}
\setcounter{page}{1}
\makeatletter
\renewcommand{\theequation}{S\arabic{equation}}
\renewcommand{\figurename}{Supplemental Material Fig.}

\section{Explicit scattering approach expressions for the current correlators}
\noindent
For the system of Fig.~1 of the main text, with arbitrary biased N-terminals ($V_1$ and $V_2$), using the general expression of~\cite{Anantram1996}, for the auto-correlator we obtain
\begin{equation*}
S_{22}/(2e^2/h)=
\begin{cases}
\left(\left|V_1\right|-\left|V_2\right|\right)B_1+\left|V_2\right|B_{+},\quad \left|V_1\right|>\left|V_2\right|,\,\text{sgn}\left(V_1V_2\right)=1, \\[8pt]
\left(\left|V_1\right|-\left|V_2\right|\right)B_1+\left|V_2\right|B_{-}, \quad \left|V_1\right|>\left|V_2\right|,\,\text{sgn}\left(V_1V_2\right)=-1, \\[8pt] 
\left(\left|V_2\right|-\left|V_1\right|\right)B_2+\left|V_1\right|B_{+}, \quad \left|V_1\right|<\left|V_2\right|,\,\text{sgn}\left(V_1V_2\right)=1, \\[8pt] 
\left(\left|V_2\right|-\left|V_1\right|\right)B_2+\left|V_1\right|B_{-}, \quad \left|V_1\right|<\left|V_2\right|,\,\text{sgn}\left(V_1V_2\right)=-1,
\end{cases}
\end{equation*}
where
\begin{equation*}
B_i=\sum\limits_{\alpha\beta\delta}\left(\frac12-\text{sgn}(\alpha\beta)T_{2i}^{\alpha\delta}\right)T_{2i}^{\beta\delta},
\end{equation*}
\begin{equation*}
B_{+}=2\sum\limits_{ij\alpha}T_{2i}^{\alpha\alpha}T_{2j}^{\alpha\overline{\alpha}}+4\left|\twoprod{2}{1}{e}{e}{2}{1}{h}{e}+\twoprod{2}{2}{e}{e}{2}{2}{h}{e}\right|^2,
\end{equation*}
\begin{equation*}
B_{-}=2\sum\limits_{i\alpha}T_{2i}^{\alpha e}T_{2i}^{\alpha h}+2\sum\limits_{\alpha\beta}T_{22}^{\alpha \beta}T_{21}^{\alpha \beta}+4\left|\twoprod{2}{1}{e}{e}{2}{1}{h}{e}+\twoprod{2}{2}{e}{h}{2}{2}{h}{h}\right|^2.
\end{equation*}
The expression for~${\cal S}_{11}$ may be obtained by the indices swap~$1\leftrightarrow2$.

\vspace{5mm}
\noindent
At the same time, the cross-correlator is
\begin{equation*}
S_{12}/(2e^2/h)=
\begin{cases}
\left(\left|V_1\right|-\left|V_2\right|\right)A_1+\left|V_2\right|A_{+},\quad \left|V_1\right|>\left|V_2\right|,\,\text{sgn}\left(V_1V_2\right)=1, \\[8pt]
\left(\left|V_1\right|-\left|V_2\right|\right)A_1+\left|V_2\right|A_{-}, \quad \left|V_1\right|>\left|V_2\right|,\,\text{sgn}\left(V_1V_2\right)=-1, \\[8pt] 
\left(\left|V_2\right|-\left|V_1\right|\right)A_2+\left|V_1\right|A_{+}, \quad \left|V_1\right|<\left|V_2\right|,\,\text{sgn}\left(V_1V_2\right)=1, \\[8pt] 
\left(\left|V_2\right|-\left|V_1\right|\right)A_2+\left|V_1\right|A_{-}, \quad \left|V_1\right|<\left|V_2\right|,\,\text{sgn}\left(V_1V_2\right)=-1,
\end{cases}
\end{equation*}
where
\begin{equation*}
A_i=-\sum\limits_{\alpha\beta\delta}\text{sgn}(\alpha\beta)T_{1i}^{\alpha\delta}T_{2i}^{\beta\delta},
\end{equation*}
\begin{equation*}
A_{+}=-\sum\limits_{\alpha\beta\delta}\text{sgn}(\alpha\beta)\left|\twoprod{1}{1}{\alpha}{\delta}{2}{1}{\beta}{\delta}+\twoprod{1}{2}{\alpha}{\delta}{2}{2}{\beta}{\delta}\right|^2,
\end{equation*}
\begin{equation*}
A_{-}=-\sum\limits_{\alpha\beta\delta}\text{sgn}(\alpha\beta)\left|\twoprod{1}{1}{\alpha}{\delta}{2}{1}{\beta}{\delta}+\twoprod{1}{2}{\alpha}{\overline{\delta}}{2}{2}{\beta}{\overline{\delta}}\right|^2.
\end{equation*}
Overall, in the case of $V_1=V_2=V>0$, for the dimensionless correlators we obtain:
\begin{equation*}
\begin{cases}
{\cal S}_{11} &=2\sum\limits_{ij\alpha}T_{1i}^{\alpha\alpha}T_{1j}^{\alpha\overline{\alpha}}+4\left|\twoprod{1}{1}{e}{e}{1}{1}{h}{e}+\twoprod{1}{2}{e}{e}{1}{2}{h}{e}\right|^2, \\[8pt]
{\cal S}_{22} &=2\sum\limits_{ij\alpha}T_{2i}^{\alpha\alpha}T_{2j}^{\alpha\overline{\alpha}}+4\left|\twoprod{2}{1}{e}{e}{2}{1}{h}{e}+\twoprod{2}{2}{e}{e}{2}{2}{h}{e}\right|^2, \\[8pt]
{\cal S}_{12} &=-\sum\limits_{\alpha\beta\delta}\text{sgn}({\alpha}\beta)|(s_{11}^{\alpha\delta})^{*}s_{21}^{\beta{\delta}}+(s_{12}^{\alpha{\delta}})^{*}s_{22}^{\beta{\delta}}|^2.
\end{cases}
\end{equation*}
\clearpage

\section{Zero-energy auto-correlators transformation}
\noindent
From now on we concentrate on the case of $V_1=V_2=V>0$. Consider, e.g., ${\cal S}_{22}$. In the zero-energy approximation, the first term is transformed as follows:
\begin{equation*}
2\sum\limits_{ij\alpha}T_{2i}^{\alpha\alpha}T_{2j}^{\alpha\overline{\alpha}}=
4\sum\limits_{j}T_{2j}^{eh}\sum\limits_{i}T_{2i}^{ee}={\cal I}_2(2-{\cal I}_2),
\end{equation*}
where we used ${\cal I}_2=2\left(T_{21}^{eh}+T_{22}^{eh}\right)$ and the sum rule.
The important corollary is
\begin{equation}
{\cal S}_{ii}\geq{\cal I}_i(2-{\cal I}_i).
\label{corrol}
\end{equation}
At the same time, using
\begin{equation*}
\left|\twoprodpr{2}{2}{e}{e}{2}{1}{h}{e}-\twoprodpr{2}{2}{h}{e}{2}{1}{e}{e}\right|^2=T_{22}^{ee}T_{21}^{he}+T_{22}^{he}T_{21}^{ee}-\twoprod{2}{1}{e}{e}{2}{1}{h}{e}\twoprod{2}{2}{h}{e}{2}{2}{e}{e}-\twoprod{2}{1}{h}{e}{2}{1}{e}{e}\twoprod{2}{2}{e}{e}{2}{2}{h}{e},
\end{equation*}
we obtain
\begin{equation*}
4\left|\twoprod{2}{1}{e}{e}{2}{1}{h}{e}+\twoprod{2}{2}{e}{e}{2}{2}{h}{e}\right|^2=4\Big\{T_{21}^{ee}T_{21}^{eh}+T_{22}^{ee}T_{22}^{eh}+\twoprod{2}{1}{e}{e}{2}{1}{h}{e}\twoprod{2}{2}{h}{e}{2}{2}{e}{e}+\twoprod{2}{1}{h}{e}{2}{1}{e}{e}\twoprod{2}{2}{e}{e}{2}{2}{h}{e}\Big\}=
\end{equation*}
\begin{equation*}
=4\Big\{T_{21}^{ee}T_{21}^{eh}+T_{22}^{ee}T_{22}^{eh}+T_{22}^{ee}T_{21}^{he}+T_{22}^{he}T_{21}^{ee}-\left|\twoprodpr{2}{2}{e}{e}{2}{1}{h}{e}-\twoprodpr{2}{2}{h}{e}{2}{1}{e}{e}\right|^2\Big\}=
\end{equation*}
\begin{equation*}
=4\left(T_{21}^{ee}+T_{22}^{ee}\right)\left(T_{21}^{eh}+T_{22}^{eh}\right)-4\left|\twoprodpr{2}{2}{e}{e}{2}{1}{h}{e}-\twoprodpr{2}{2}{h}{e}{2}{1}{e}{e}\right|^2={\cal I}_2(2-{\cal I}_2)-4\left|\twoprodpr{2}{2}{e}{e}{2}{1}{h}{e}-\twoprodpr{2}{2}{h}{e}{2}{1}{e}{e}\right|^2.
\end{equation*}
Overall, the auto-correlators may be rewritten as follows:
\begin{equation*}
\begin{cases}
{\cal S}_{11}=2{\cal I}_1(2-{\cal I}_1)-4\left|\twoprodpr{1}{1}{e}{e}{1}{2}{h}{e}-\twoprodpr{1}{1}{h}{e}{1}{2}{e}{e}\right|^2, \\[10pt]
{\cal S}_{22}=2{\cal I}_2(2-{\cal I}_2)-4\left|\twoprodpr{2}{2}{e}{e}{2}{1}{h}{e}-\twoprodpr{2}{2}{h}{e}{2}{1}{e}{e}\right|^2,
\end{cases}
\end{equation*}
or, written with the use of $p_{11}$,
\begin{equation}
\begin{cases}
{\cal S}_{11}=2{\cal I}_1(2-{\cal I}_1)-2p_{11}, \\[10pt]
{\cal S}_{22}=2{\cal I}_2(2-{\cal I}_2)-2p_{11}.
\end{cases}
\label{expr_for_b}
\end{equation}
\clearpage

\section{Zero-energy cross-correlator transformation}
\noindent
Consider the orthogonality rule
\begin{equation*}
\sum\limits_{l\delta}(s_{il}^{\alpha\delta})^{\dagger}s_{jl}^{\beta\delta}=\delta_{ij}\delta_{\alpha\beta}
\end{equation*}
for the particular case of $i=1$, $j=2$. We designate

\begin{center}
\vspace{3mm}
\textcolor{red}{$\alpha=1$, $\beta=1$:}

$a=(s_{11}^{ee})^{\dagger}s_{21}^{ee}$, $b=(s_{11}^{eh})^{\dagger}s_{21}^{eh}$, $c=(s_{12}^{ee})^{\dagger}s_{22}^{ee}$, $d=(s_{12}^{eh})^{\dagger}s_{22}^{eh}$.

\vspace{3mm}
\textcolor{red}{$\alpha=1$, $\beta=2$:}

$e=(s_{11}^{ee})^{\dagger}s_{21}^{he}$, $f=(s_{11}^{eh})^{\dagger}s_{21}^{hh}$, $g=(s_{12}^{ee})^{\dagger}s_{22}^{he}$, $h=(s_{12}^{eh})^{\dagger}s_{22}^{hh}$.

\vspace{3mm}
\textcolor{red}{$\alpha=2$, $\beta=1$:}

$k=(s_{11}^{he})^{\dagger}s_{21}^{ee}$, $l=(s_{11}^{hh})^{\dagger}s_{21}^{eh}$, $m=(s_{12}^{he})^{\dagger}s_{22}^{ee}$, $n=(s_{12}^{hh})^{\dagger}s_{22}^{eh}$.

\vspace{3mm}
\textcolor{red}{$\alpha=2$, $\beta=2$:}

$p=(s_{11}^{he})^{\dagger}s_{21}^{he}$, $q=(s_{11}^{hh})^{\dagger}s_{21}^{hh}$, $r=(s_{12}^{he})^{\dagger}s_{22}^{he}$, $s=(s_{12}^{hh})^{\dagger}s_{22}^{hh}$.
\end{center}

\vspace{5mm}
\noindent
The current cross-correlator is
\begin{equation*}
{\cal S}_{12}=-\sum\limits_{\alpha\beta\delta}\text{sgn}({\alpha}\beta)|(s_{11}^{\alpha\delta})^{*}s_{21}^{\beta{\delta}}+(s_{12}^{\alpha{\delta}})^{*}s_{22}^{\beta{\delta}}|^2=
\end{equation*}
\begin{equation*}
=\left\{|e+g|^2+|f+h|^2+|k+m|^2+|l+n|^2\right\}-\left\{|a+c|^2+|b+d|^2+|p+r|^2+|q+s|^2\right\}=
\end{equation*}
\begin{equation*}
=E+G+F+H+K+M+L+N+\left\{eg^*+fh^*+km^*+ln^*+\text{c.c.}\right\}-
\end{equation*}
\begin{equation*}
-(A+C+B+D+P+R+Q+S)-\left\{ac^*+bd^*+pr^*+qs^*+\text{c.c.}\right\}.
\end{equation*}

\vspace{1mm}
\begin{equation*}
ac^*-eg^*=(s_{11}^{ee})^{\dagger}s_{12}^{ee}\textcolor{red}{\left\{(s_{22}^{ee})^{\dagger}s_{21}^{ee}-(s_{22}^{he})^{\dagger}s_{21}^{he}\right\}},\quad bd^*-fh^*=(s_{11}^{eh})^{\dagger}s_{12}^{eh}\textcolor{blue}{\left\{(s_{22}^{eh})^{\dagger}s_{21}^{eh}-(s_{22}^{hh})^{\dagger}s_{21}^{hh}\right\}},
\end{equation*}
\begin{equation*}
pr^*-km^*=(s_{11}^{he})^{\dagger}s_{12}^{he}\textcolor{red}{\left\{(s_{22}^{he})^{\dagger}s_{21}^{he}-(s_{22}^{ee})^{\dagger}s_{21}^{ee}\right\}},\quad qs^*-ln^*=(s_{11}^{hh})^{\dagger}s_{12}^{hh}\textcolor{blue}{\left\{(s_{22}^{hh})^{\dagger}s_{21}^{hh}-(s_{22}^{eh})^{\dagger}s_{21}^{eh}\right\}}.
\end{equation*}

\vspace{1mm}
\noindent
The curly brackets give
\begin{equation*}
=-\textcolor{red}{\left\{(s_{22}^{ee})^{\dagger}s_{21}^{ee}-(s_{22}^{he})^{\dagger}s_{21}^{he}\right\}}\left\{(s_{11}^{ee})^{\dagger}s_{12}^{ee}-(s_{11}^{he})^{\dagger}s_{12}^{he}\right\}-\textcolor{blue}{\left\{(s_{22}^{hh})^{\dagger}s_{21}^{hh}-(s_{22}^{eh})^{\dagger}s_{21}^{eh}\right\}}\left\{(s_{11}^{hh})^{\dagger}s_{12}^{hh}-(s_{11}^{eh})^{\dagger}s_{12}^{eh}\right\}+\text{c.c}=
\end{equation*}
\begin{equation*}
=-\left\{xy^*+uv^*+x^*y+u^*v\right\}=-|x+y|^2+X+Y-|u+v|^2+U+V,
\end{equation*}
where
\begin{equation*}
x=(s_{22}^{ee})^{\dagger}s_{21}^{ee}-(s_{22}^{he})^{\dagger}s_{21}^{he},\quad y^*=(s_{11}^{ee})^{\dagger}s_{12}^{ee}-(s_{11}^{he})^{\dagger}s_{12}^{he},
\end{equation*}
\begin{equation*}
u=(s_{22}^{hh})^{\dagger}s_{21}^{hh}-(s_{22}^{eh})^{\dagger}s_{21}^{eh},\quad v^*=(s_{11}^{hh})^{\dagger}s_{12}^{hh}-(s_{11}^{eh})^{\dagger}s_{12}^{eh},
\end{equation*}
\begin{equation*}
X=\left|(s_{22}^{ee})^{\dagger}s_{21}^{ee}-(s_{22}^{he})^{\dagger}s_{21}^{he}\right|^2=T_{22}^{ee}T_{21}^{ee}+T_{22}^{he}T_{21}^{he}-\left\{(s_{22}^{ee})^{\dagger}s_{21}^{ee}s_{22}^{he}(s_{21}^{he})^{\dagger}+\text{c.c.}\right\}=
\end{equation*}
\begin{equation*}
=T_{22}^{ee}T_{21}^{ee}+T_{22}^{he}T_{21}^{he}+\left|s_{22}^{ee}s_{21}^{he}-s_{22}^{he}s_{21}^{ee}\right|^2-T_{22}^{ee}T_{21}^{he}-T_{22}^{he}T_{21}^{ee}=\left(T_{22}^{ee}-T_{22}^{he}\right)\left(T_{21}^{ee}-T_{21}^{he}\right)+\frac{p_{11}}{2},
\end{equation*}
and similarly,
\begin{equation*}
Y=\left(T_{11}^{ee}-T_{11}^{he}\right)\left(T_{12}^{ee}-T_{12}^{he}\right)+\frac{p_{11}}{2}, \quad U=X,\quad V=Y.
\end{equation*}
Therefore
\begin{equation*}
{\cal S}_{12}=E+F+G+H+K+L+M+N-(A+B+C+D+P+Q+R+S)+
\end{equation*}
\begin{equation*}
+X+Y+U+V-|x+y|^2-|u+v|^2=
\end{equation*}
\begin{equation*}
=(E+F+G+H-A-B-C-D)+(K+L+M+N-P-Q-R-S)+
\end{equation*}
\begin{equation*}
+X+Y+U+V-|x+y|^2-|u+v|^2=
\end{equation*}
\begin{equation*}
=T_{11}^{ee}T_{21}^{he}+T_{11}^{eh}T_{21}^{hh}+T_{12}^{ee}T_{22}^{he}+T_{12}^{eh}T_{22}^{hh}-T_{11}^{ee}T_{21}^{ee}-T_{11}^{eh}T_{21}^{eh}-T_{12}^{ee}T_{22}^{ee}-T_{12}^{eh}T_{22}^{eh}+
\end{equation*}
\begin{equation*}
+T_{11}^{he}T_{21}^{ee}+T_{11}^{hh}T_{21}^{eh}+T_{12}^{he}T_{22}^{ee}+T_{12}^{hh}T_{22}^{eh}-T_{11}^{he}T_{21}^{he}-T_{11}^{hh}T_{21}^{hh}-T_{12}^{he}T_{22}^{he}-T_{12}^{hh}T_{22}^{hh}+
\end{equation*}
\begin{equation*}
+2p_{11}+2\left(T_{11}^{ee}-T_{11}^{he}\right)\left(T_{12}^{ee}-T_{12}^{he}\right)+2\left(T_{22}^{ee}-T_{22}^{eh}\right)\left(T_{21}^{ee}-T_{21}^{eh}\right)-|x+y|^2-|u+v|^2=
\end{equation*}
\begin{equation*}
=2T_{11}^{ee}\left(T_{21}^{eh}-T_{21}^{ee}+T_{12}^{ee}-T_{12}^{eh}\right)+2T_{11}^{eh}\left(T_{21}^{ee}-T_{21}^{eh}+T_{12}^{eh}-T_{12}^{ee}\right)+
\end{equation*}
\begin{equation*}
+2T_{22}^{eh}\left(T_{12}^{ee}-T_{12}^{eh}-T_{21}^{ee}+T_{21}^{eh}\right)+2T_{22}^{ee}\left(T_{12}^{eh}-T_{12}^{ee}+T_{21}^{ee}-T_{21}^{eh}\right)+
\end{equation*}
\begin{equation*}
+2p_{11}-|x+y|^2-|u+v|^2=
\end{equation*}
\begin{equation*}
=2\left\{T_{12}^{eh}-T_{12}^{ee}+T_{21}^{ee}-T_{21}^{eh}\right\}\left\{T_{22}^{ee}-T_{22}^{eh}+T_{11}^{eh}-T_{11}^{ee}\right\}+
\end{equation*}
\begin{equation*}
+2p_{11}-4\left|(s_{12}^{ee})^{\dagger}s_{11}^{ee}+(s_{22}^{ee})^{\dagger}s_{21}^{ee}\right|^2-4\left|(s_{12}^{hh})^{\dagger}s_{11}^{hh}+(s_{22}^{hh})^{\dagger}s_{21}^{hh}\right|^2=
\end{equation*}
\begin{equation*}
\textcolor{gray}{T_{21}^{ee}-T_{21}^{eh}+T_{12}^{eh}-T_{12}^{ee}=1-T_{22}^{ee}-T_{22}^{he}-2T_{21}^{eh}+T_{12}^{eh}-T_{12}^{ee}=2\left(T_{12}^{he}-T_{21}^{eh}\right)}.
\end{equation*}
\begin{equation*}
=2p_{11}+8\left(T_{12}^{eh}-T_{21}^{eh}\right)\left(T_{11}^{eh}-T_{22}^{eh}\right)-8\left|(s_{12}^{ee})^{\dagger}s_{11}^{ee}+(s_{22}^{ee})^{\dagger}s_{21}^{ee}\right|^2=
\end{equation*}
\begin{equation*}
=2p_{11}+8\left(T_{12}^{eh}-T_{21}^{eh}\right)\left(T_{11}^{eh}-T_{22}^{eh}\right)-8\left|(s_{12}^{he})^{\dagger}s_{11}^{he}+(s_{22}^{he})^{\dagger}s_{21}^{he}\right|^2.
\end{equation*}

\vspace{3mm}
\noindent
Consider
\begin{equation*}
\left|(s_{12}^{he})^{\dagger}s_{11}^{he}+(s_{22}^{he})^{\dagger}s_{21}^{he}\right|^2=T_{12}^{he}T_{11}^{he}+T_{22}^{he}T_{21}^{he}+(s_{12}^{he})^{\dagger}s_{11}^{he}(s_{21}^{he})^{\dagger}s_{22}^{he}+\text{c.c.}=
\end{equation*}
\begin{equation*}
=\left(T_{12}^{he}+T_{22}^{he}\right)\left(T_{11}^{he}+T_{21}^{he}\right)-\left|s_{12}^{he}s_{21}^{he}-s_{11}^{he}s_{22}^{he}\right|^2.
\end{equation*}
Therefore,
\begin{equation*}
{\cal S}_{12}=2p_{11}+8\left(T_{12}^{he}-T_{21}^{he}\right)\left(T_{11}^{he}-T_{22}^{he}\right)-8\left(T_{12}^{he}+T_{22}^{he}\right)\left(T_{11}^{he}+T_{21}^{he}\right)+8\left|s_{12}^{he}s_{21}^{he}-s_{11}^{he}s_{22}^{he}\right|^2=
\end{equation*}
\begin{equation*}
=2p_{11}-2{\cal I}_1{\cal I}_2+8\left|s_{12}^{he}s_{21}^{he}-s_{11}^{he}s_{22}^{he}\right|^2.
\end{equation*}
\clearpage

\section{Some boundaries on splitting probability and efficiency}
\noindent
Together, (\ref{corrol}) and (\ref{expr_for_b}) imply
\begin{equation*}
p_{11}={\cal I}_1(2-{\cal I}_1)-\frac{{\cal S}_{11}}{2}\leq\frac{{\cal I}_1(2-{\cal I}_1)}{2}={\cal I}_1-\frac{{\cal I}_1^2}{2}\leq\frac12
\end{equation*}
and, similarly
\begin{equation*}
p_{11}\leq{\cal I}_2-\frac{{\cal I}_2^2}{2}\leq\frac12.
\end{equation*}
The sum of two last inequalities gives
\begin{equation*}
2p_{11}\leq{\cal I}_1+{\cal I}_2-\frac{{\cal I}_1^2+{\cal I}_2^2}{2}\leq{\cal I}_1+{\cal I}_2-\frac{({\cal I}_1+{\cal I}_2)^2}{4},
\end{equation*}
so that the splitting efficiency is
\begin{equation*}
K=\frac{2p_{11}}{{\cal I}_1+{\cal I}_2}\leq1-\frac{|{\cal I}_3|}{4}.
\end{equation*}

To find when $p_{11}$ achieves $1/2$ in terms of scattering matrix, we start with the following expression for~$p_{11}$:
\begin{equation*}
p_{11}=2\left|\twoprodpr{1}{1}{e}{e}{1}{2}{h}{e}-\twoprodpr{1}{1}{h}{e}{1}{2}{e}{e}\right|^2.
\end{equation*}
For brevity we introduce
\begin{equation*}
a=s_{11}^{ee},\quad b=s_{12}^{he},\quad c=s_{11}^{he},\quad d=s_{12}^{ee}
\end{equation*}
and
\begin{equation*}
A=|a|^2,\quad B=|b|^2,\quad C=|c|^2,\quad D=|d|^2.
\end{equation*}
At zero energy the sum rule implies $A+B+C+D=1$.
We also introduce
\begin{equation*}
K=abc^{*}d^{*}+a^{*}b^{*}cd,
\end{equation*}
\begin{equation*}
k_1=|ac^{*}+b^{*}d|^2=AC+BD+K,
\end{equation*}
\begin{equation*}
k_2=|ad^{*}+b^{*}c|^2=AD+BC+K.
\end{equation*}
From here
$K=k_1-AC-BD=k_2-AD-BC$. Then
\begin{equation*}
p_{11}=2|ab-cd|^2=2\left\{|ab|^2+|cd|^2-\left(abc^{*}d^{*}+a^{*}b^{*}cd\right)\right\}=
\end{equation*}
\begin{equation*}
=2AB+2CD+AC+BD-k_1+AD+BC-k_2=2AB+2CD+(A+B)(C+D)-k_1-k_2=
\end{equation*}
\begin{equation*}
=(A+B)^2-A^2-B^2+(C+D)^2-C^2-D^2+(A+B)(C+D)-k_1-k_2=
\end{equation*}
\begin{equation*}
=\frac12(A+B)^2+\frac12(C+D)^2+(A+B)(C+D)-k_1-k_2+\frac12(A+B)^2+\frac12(C+D)^2-A^2-B^2-C^2-D^2=
\end{equation*}
\begin{equation*}
=\frac12-k_1-k_2-\frac12(A-B)^2-\frac12(C-D)^2.
\end{equation*}
Therefore, $p_{11}\leq1/2$ and to achieve this value one needs to zero all four subtrahends in the last expression for~$p_{11}$.
\clearpage

\section{$s$-matrix of a single-mode NSN device derived from its NNN counterpart}
\begin{figure}[h]
\begin{center}
\includegraphics[width=\linewidth]{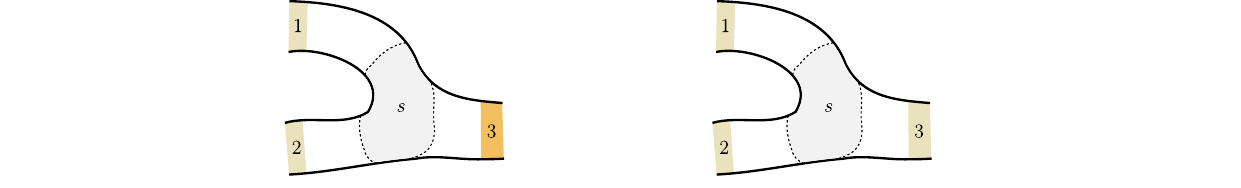}
\end{center}
\caption{(Left)~Sketch of the NSN device. The S-terminal is assumed ideal, producing only Andreev scattering. Any interface scattering belongs to the scattering region inside the conductor, marked by gray color. (Right)~Sketch of the NNN~device with the S-terminal substituted by the N-terminal.
}
\label{fig_identity}
\end{figure}

The $s$-matrix of the original NSN device is denoted $s_{ij}^{\alpha\beta}$ as in the main text, the $s$-matrix of the NNN device is denoted $\tilde{s}_{ij}^{\alpha\alpha}$. The Andreev scattering amplitude from the S-terminal is denoted by $A^{he} = -(A^{eh})^*$ and $|A^{he}|=1$ by assumption.   

Consider the process of local Andreev reflection (LAR). The amplitude of this process is obtained by a sum of all possible scattering amplitudes which involve odd number of~ARs from the S-terminal. Therefore:

\begin{equation*}
s_{11}^{he} = \tilde{s}_{13}^{hh}A^{he}\tilde{s}_{31}^{ee} + \tilde{s}_{13}^{hh}A^{he}\left(\tilde{s}_{33}^{ee}A^{eh}\tilde{s}_{33}^{hh}A^{he}\right)\tilde{s}_{31}^{ee}+\tilde{s}_{13}^{hh}A^{he}\left(\tilde{s}_{33}^{ee}A^{eh}\tilde{s}_{33}^{hh}A^{he}\right)^2\tilde{s}_{31}^{ee}+\ldots
\end{equation*}

\noindent
The first term in this expression is interpreted as follows. The electron from the N-terminal 1  reaches the S-terminal with the same amplitude as the electron reaches the N-terminal 3 in the corresponding NNN device ($\tilde{s}_{31}^{ee}$). After that, the electron experiences a single AR with the amplitude of $A^{he}$ and goes back to the N-terminal 1 as a hole ($\tilde{s}_{13}^{hh}$). The second term is different in that the quasiparticle experiences several Andreev reflection from the S-terminal, which become possible owing to back-scattering amplitudes $\tilde{s}_{33}^{ee}$ and $\tilde{s}_{33}^{hh}$ of the electron and hole in the corresponding NNN~device. The sum of the above series gives the finite expression:
\begin{align}
	s_{11}^{he} &= \tilde{s}_{13}^{hh}\tilde{s}_{31}^{ee}A^{he}(1+\tilde{s}_{33}^{ee}\tilde{s}_{33}^{hh})^{-1}.\label{s11he}
\end{align}
Similarly, one obtains three other Andreev amplitudes:
\begin{align}
	s_{22}^{he} &= \tilde{s}_{23}^{hh}\tilde{s}_{32}^{ee}A^{he}(1+\tilde{s}_{33}^{ee}\tilde{s}_{33}^{hh})^{-1},\label{s22he} \\
		s_{12}^{he} &= \tilde{s}_{13}^{hh}\tilde{s}_{32}^{ee}A^{he}(1+\tilde{s}_{33}^{ee}\tilde{s}_{33}^{hh})^{-1},\label{s12he} \\
		s_{21}^{he} &= \tilde{s}_{23}^{hh}\tilde{s}_{31}^{ee}A^{he}(1+\tilde{s}_{33}^{ee}\tilde{s}_{33}^{hh})^{-1}.\label{s21he} 
\end{align}
Obviously, the eqs.~(\ref{s11he}-\ref{s21he}) prove the identity in the NSN device $s_{11}^{he}s_{22}^{he} =s_{12}^{he}s_{21}^{he}$.

\vspace{3mm}
The derivation of normal scattering amplitudes is similar. Here one takes into account all the amplitudes with even number of AR. For example, for the backscattering amplitude $s_{11}^{ee}$:
\begin{align*}
s_{11}^{ee} &= \tilde{s}_{11}^{ee}+ \tilde{s}_{13}^{ee}A^{eh}\tilde{s}_{33}^{hh}A^{he}\tilde{s}_{31}^{ee}+\tilde{s}_{13}^{ee}A^{eh}\tilde{s}_{33}^{hh}A^{he}\left(\tilde{s}_{33}^{ee}A^{eh}\tilde{s}_{33}^{hh}A^{he}\right)\tilde{s}_{31}^{ee}+\tilde{s}_{13}^{ee}A^{eh}\tilde{s}_{33}^{hh}A^{he}\left(\tilde{s}_{33}^{ee}A^{eh}\tilde{s}_{33}^{hh}A^{he}\right)^2\tilde{s}_{31}^{ee}+\ldots
\end{align*}
The sum of this series gives:
\begin{align*}
	s_{11}^{ee} = \tilde{s}_{11}^{ee}-\tilde{s}_{13}^{ee}\tilde{s}_{33}^{hh}\tilde{s}_{31}^{ee}(1+\tilde{s}_{33}^{ee}\tilde{s}_{33}^{hh})^{-1}.
\end{align*}
and similarly for the other three normal amplitudes in the NSN devices:
\begin{align*}
	s_{22}^{ee} &= \tilde{s}_{22}^{ee}-\tilde{s}_{23}^{ee}\tilde{s}_{33}^{hh}\tilde{s}_{32}^{ee}(1+\tilde{s}_{33}^{ee}\tilde{s}_{33}^{hh})^{-1},\\
		s_{12}^{ee} &= \tilde{s}_{12}^{ee}-\tilde{s}_{13}^{ee}\tilde{s}_{33}^{hh}\tilde{s}_{32}^{ee}(1+\tilde{s}_{33}^{ee}\tilde{s}_{33}^{hh})^{-1},\\
			s_{21}^{ee} &= \tilde{s}_{21}^{ee}-\tilde{s}_{23}^{ee}\tilde{s}_{33}^{hh}\tilde{s}_{31}^{ee}(1+\tilde{s}_{33}^{ee}\tilde{s}_{33}^{hh})^{-1}.
\end{align*}
The missing terms of the s-matrix are obtained from these expression by general symmetries $s_{ij}^{eh}=-(s_{ij}^{eh})^*$ and $s_{ij}^{ee}=(s_{ij}^{hh})^*$. 
\clearpage

\section{Splitting probability and efficiency in a single-mode NSN device}
\noindent
We will further use the following identity:
\begin{equation*}
s_{21}^{he}s_{12}^{he}=s_{11}^{he}s_{22}^{he}\quad\to\quad 
\end{equation*}
\begin{equation*}
\quad\to\quad s_{21}^{he}s_{12}^{he}\left(s_{21}^{he}\right)^*\left(s_{11}^{he}\right)^*=s_{11}^{he}s_{22}^{he}\left(s_{21}^{he}\right)^*\left(s_{11}^{he}\right)^*,
\end{equation*}
\begin{equation*}
s_{21}^{he}s_{12}^{he}\left(s_{21}^{he}\right)^*\left(s_{11}^{he}\right)^*=s_{11}^{he}s_{22}^{he}\left(s_{21}^{he}\right)^*\left(s_{11}^{he}\right)^*,
\end{equation*}
\begin{equation}
T_{21}^{he}s_{12}^{he}\left(s_{11}^{he}\right)^*=T_{11}^{he}s_{22}^{he}\left(s_{21}^{he}\right)^*.
\label{usid}
\end{equation}
We start with the two equivalent expressions for the splitting probability:
\begin{equation*}
p_{11}=2\left|\twoprodpr{1}{1}{e}{e}{1}{2}{h}{e}-\twoprodpr{1}{1}{h}{e}{1}{2}{e}{e}\right|^2=2\left|\twoprodpr{2}{2}{e}{e}{2}{1}{h}{e}-\twoprodpr{2}{2}{h}{e}{2}{1}{e}{e}\right|^2.
\end{equation*}
It is now convenient to write
\begin{equation*}
p_{11}=2\Bigg\{\frac{T_{21}^{he}}{T_{21}^{he}+T_{11}^{he}}\left|\twoprodpr{1}{1}{e}{e}{1}{2}{h}{e}-\twoprodpr{1}{1}{h}{e}{1}{2}{e}{e}\right|^2+\frac{T_{11}^{he}}{T_{21}^{he}+T_{11}^{he}}\left|\twoprodpr{2}{2}{e}{e}{2}{1}{h}{e}-\twoprodpr{2}{2}{h}{e}{2}{1}{e}{e}\right|^2\Bigg\}=
\end{equation*}
\begin{equation*}
=\frac{2}{T_{21}^{he}+T_{11}^{he}}\Bigg\{T_{21}^{he}\left|\twoprodpr{1}{1}{e}{e}{1}{2}{h}{e}-\twoprodpr{1}{1}{h}{e}{1}{2}{e}{e}\right|^2+T_{11}^{he}\left|\twoprodpr{2}{2}{e}{e}{2}{1}{h}{e}-\twoprodpr{2}{2}{h}{e}{2}{1}{e}{e}\right|^2\Bigg\}=
\end{equation*}
\begin{equation*}
=\frac{2}{T_{21}^{he}+T_{11}^{he}}\Bigg\{T_{21}^{he}\Big[T_{11}^{ee}T_{12}^{he}+T_{11}^{he}T_{12}^{ee}-\big(s_{11}^{ee}s_{12}^{he}(s_{11}^{he})^*(s_{12}^{ee})^*+\text{c.c.}\big)\Big]+
\end{equation*}
\begin{equation*}
+T_{11}^{he}\Big[T_{22}^{ee}T_{21}^{he}+T_{22}^{he}T_{21}^{ee}-\big((s_{22}^{ee})^*(s_{21}^{he})^*s_{22}^{he}s_{21}^{ee}+\text{c.c.}\big)\Big]\Bigg\}.
\end{equation*}
Using~(\ref{usid}), we obtain
\begin{equation*}
p_{11}=\frac{2}{T_{21}^{he}+T_{11}^{he}}\Bigg\{T_{21}^{he}\Big[T_{11}^{ee}T_{12}^{he}+T_{11}^{he}T_{12}^{ee}\Big]-T_{21}^{he}\big(s_{11}^{ee}s_{12}^{he}(s_{11}^{he})^*(s_{12}^{ee})^*+\text{c.c.}\big)+
\end{equation*}
\begin{equation*}
+T_{11}^{he}\Big[T_{22}^{ee}T_{21}^{he}+T_{22}^{he}T_{21}^{ee}\Big]-T_{11}^{he}\big((s_{22}^{ee})^*(s_{21}^{he})^*s_{22}^{he}s_{21}^{ee}+\text{c.c.}\big)\Bigg\}.
\end{equation*}
\begin{equation*}
=\frac{2}{T_{21}^{he}+T_{11}^{he}}\Bigg\{T_{21}^{he}\Big[T_{11}^{ee}T_{12}^{he}+T_{11}^{he}T_{12}^{ee}\Big]-T_{11}^{he}\big(s_{11}^{ee}s_{22}^{he}(s_{21}^{he})^*(s_{12}^{ee})^*+\text{c.c.}\big)+
\end{equation*}
\begin{equation*}
+T_{11}^{he}\Big[T_{22}^{ee}T_{21}^{he}+T_{22}^{he}T_{21}^{ee}\Big]-T_{11}^{he}\big((s_{22}^{ee})^*(s_{21}^{he})^*s_{22}^{he}s_{21}^{ee}+\text{c.c.}\big)\Bigg\}=
\end{equation*}
\begin{equation*}
=\frac{2}{T_{21}^{he}+T_{11}^{he}}\Bigg\{T_{21}^{he}\Big[T_{11}^{ee}T_{12}^{he}+T_{11}^{he}T_{12}^{ee}\Big]+T_{11}^{he}\Big[T_{22}^{ee}T_{21}^{he}+T_{22}^{he}T_{21}^{ee}\Big]-T_{11}^{he}\Big((s_{21}^{he})^*s_{22}^{he}\big\{(s_{12}^{ee})^*s_{11}^{ee}+(s_{22}^{ee})^*s_{21}^{ee}\big\}+\text{c.c.}\Big)\Bigg\}=
\end{equation*}
\begin{equation*}
=\frac{2}{T_{21}^{he}+T_{11}^{he}}\Bigg\{T_{21}^{he}\Big[T_{11}^{ee}T_{12}^{he}+T_{11}^{he}T_{12}^{ee}\Big]+T_{11}^{he}\Big[T_{22}^{ee}T_{21}^{he}+T_{22}^{he}T_{21}^{ee}\Big]+T_{11}^{he}\Big((s_{21}^{he})^*s_{22}^{he}\big\{(s_{12}^{he})^*s_{11}^{he}+(s_{22}^{he})^*s_{21}^{he}\big\}+\text{c.c.}\Big)\Bigg\}=
\end{equation*}
\begin{equation*}
=\frac{2}{T_{21}^{he}+T_{11}^{he}}\Bigg\{T_{21}^{he}\Big[T_{11}^{ee}T_{12}^{he}+T_{11}^{he}T_{12}^{ee}\Big]+T_{11}^{he}\Big[T_{22}^{ee}T_{21}^{he}+T_{22}^{he}T_{21}^{ee}\Big]+T_{11}^{he}\Big((s_{21}^{he})^*s_{22}^{he}\big\{(s_{12}^{he})^*s_{11}^{he}+(s_{22}^{he})^*s_{21}^{he}\big\}+\text{c.c.}\Big)\Bigg\}=
\end{equation*}
\begin{equation*}
=\frac{2}{T_{21}^{he}+T_{11}^{he}}\Bigg\{T_{21}^{he}\Big[T_{11}^{ee}T_{12}^{he}+T_{11}^{he}T_{12}^{ee}\Big]+T_{11}^{he}\Big[T_{22}^{ee}T_{21}^{he}+T_{22}^{he}T_{21}^{ee}\Big]+2T_{11}^{he}\Big(T_{22}^{he}T_{11}^{he}+T_{22}^{he}T_{21}^{he}\Big)\Bigg\}=
\end{equation*}
\begin{equation*}
=\frac{2}{T_{21}^{he}+T_{11}^{he}}\Bigg\{T_{21}^{he}T_{12}^{he}\Big[T_{11}^{ee}+T_{21}^{ee}+T_{11}^{he}+T_{21}^{he}\Big]+T_{21}^{he}T_{11}^{he}\Big[T_{12}^{ee}+T_{22}^{ee}+T_{12}^{he}+T_{22}^{he}\Big]\Bigg\}=
\end{equation*}
\begin{equation*}
=2T_{21}^{he}\frac{T_{12}^{he}+T_{11}^{he}}{T_{21}^{he}+T_{11}^{he}}=\frac{T_{21}^{he}}{T_{21}^{he}+T_{11}^{he}}{\cal I}_1=p_{11}.
\end{equation*}
Note that this expression doesn't change with index swap so that in terms of Andreev transmission probabilities one has
\begin{equation*}
p_{11}=2T_{21}^{he}\frac{T_{12}^{he}+T_{11}^{he}}{T_{21}^{he}+T_{11}^{he}}=2T_{12}^{he}\frac{T_{21}^{he}+T_{22}^{he}}{T_{12}^{he}+T_{22}^{he}}.
\end{equation*}
We now notice that
\begin{equation*}
{\cal I}_1T_{21}^{he}=2(T_{11}^{he}+T_{12}^{he})T_{21}^{he}=2(T_{11}^{he}T_{21}^{he}+T_{12}^{he}T_{21}^{he})=2(T_{11}^{he}T_{21}^{he}+T_{11}^{he}T_{22}^{he})=2(T_{21}^{he}+T_{22}^{he})T_{11}^{he}={\cal I}_2T_{11}^{he}.
\end{equation*}
In terms of the measurable quantities one therefore obtains
\begin{equation*}
p_{11}=\frac{1}{1+T_{21}^{he}/T_{11}^{he}}{\cal I}_2=\frac{1}{1+{\cal I}_2/{\cal I}_1}{\cal I}_2=\frac{{\cal I}_1{\cal I}_2}{{\cal I}_1+{\cal I}_2}.
\end{equation*}
The splitting efficiency is then
\begin{equation*}
K=\frac{2p_{11}}{{\cal I}_1+{\cal I}_2}=\frac{2{\cal I}_1{\cal I}_2}{({\cal I}_1+{\cal I}_2)^2}.
\end{equation*}
\clearpage

\section{Beam Splitter analytical result}
Splitting probability is
\begin{equation*}
p_{11}=\frac{\tau^2\left(8Z^4\tau^2-32Z^4\tau+16Z^4+8Z^2\tau^2-16Z^2\tau+8Z^2+2\tau^2-4\tau+2\right)+8Z^2\tau^2\sqrt{1-2\tau}(2Z^2+1)(1-\tau)}{4\left(4Z^4\tau^2+4Z^2\tau^2-8Z^2\tau+4Z^2+\tau^2-2\tau+1\right)^2}
\end{equation*}
The average currents and current correlators are defined via
\begin{equation*}
\begin{cases}
{\cal I}_{1}={\cal I}_{2}=2p_{11}, \\[8pt] {\cal S}_{11}=2{\cal I}_1(2-{\cal I}_1)-2p_{11}, \\[8pt] {\cal S}_{12}=2p_{11}-2{\cal I}_1^2.
\end{cases}
\end{equation*}
\clearpage

\section{HBI scattering matrix}
Using the notations of~\cite{Khrapai2023}, for the scattering matrix elements of the BHI we obtain
\begin{equation*}
\Phi=\phi_A+\phi_B+\phi_C+\phi_{r_1'}+\phi_{r_2}+\phi_{s_1},
\end{equation*}
\begin{equation*}
\Phi_1=-\phi_A+\phi_B+\phi_{t_1'}+\phi_{r_2}+\phi_{s_2}.
\end{equation*}
\vspace{3mm}
\begin{equation*}
s_{11}^{ee(hh)}=\frac{1}{\sqrt{R_1}}\left[1-\frac{T_1\left(1-\sqrt{R_1R_2}\left|t_{ee}\right|e^{\mp i\Phi}\right)}{\dnmrt}\right]e^{\pm i\phi_{r_1}},
\end{equation*}
\vspace{3mm}
\begin{equation*}
s_{11}^{eh(he)}=\pm\frac{\sqrt{R_2}T_1\left|t_{eh}\right|}{\dnmrt}e^{\pm i\left(\Phi_1-\phi_{t_1}+\phi_C\right)},
\end{equation*}
\vspace{3mm}
\begin{equation*}
s_{21}^{ee(hh)}=-\frac{\sqrt{T_1T_2}\left(\sqrt{R_1R_2}-\left|t_{ee}\right|e^{\pm i\Phi}\right)}{\dnmrt}e^{\pm i\left(\phi_{t_1}+\phi_{t_2}-\phi_{C}-\phi_{r_2}-\phi_{r_1'}\right)},
\end{equation*}
\vspace{3mm}
\begin{equation*}
s_{21}^{eh(he)}=\pm\frac{\sqrt{T_1T_2}\left|t_{eh}\right|}{\dnmrt}e^{\pm i\left(-\phi_{A}+\phi_{B}+\phi_{s_2}+\phi_{t_2}-\phi_{t_1}\right)},
\end{equation*}
\vspace{3mm}
\begin{equation*}
s_{12}^{ee(hh)}=\frac{\sqrt{T_1T_2}\left(1-\sqrt{R_1R_2}\left|t_{ee}\right|e^{\mp i\Phi}\right)}{\dnmrt}e^{\pm i\left(\phi_{C}+\phi_{t_1'}+\phi_{t_2'}\right)},
\end{equation*}
\vspace{3mm}
\begin{equation*}
s_{12}^{eh(he)}=\pm\frac{\sqrt{R_1R_2T_1T_2}\left|t_{eh}\right|}{\dnmrt}e^{\pm i\left(\Phi_1-\phi_{r_1'}-\phi_{t_2'}\right)},
\end{equation*}
\vspace{3mm}
\begin{equation*}
s_{22}^{ee(hh)}=\frac{1}{\sqrt{R_2}}\left[1-\frac{T_2\left(1-\sqrt{R_1R_2}\left|t_{ee}\right|e^{\mp i\Phi}\right)}{\dnmrt}\right]e^{\pm i\phi_{r_2'}},
\end{equation*}
\vspace{3mm}
\begin{equation*}
s_{22}^{eh(he)}=\mp\frac{\sqrt{R_1}T_2\left|t_{eh}\right|}{\dnmrt}e^{\pm i\left(2\phi_B+\phi_{s_1}+\phi_{s_2}+2\phi_{r_2}+\phi_{r_2'}-2\phi_{t_2'}-\Phi\right)}.
\end{equation*}
\clearpage

\section{Current cross-correlator and splitting probability in HBI}
\begin{figure}[h]
\begin{center}
\includegraphics[width=.65\linewidth]{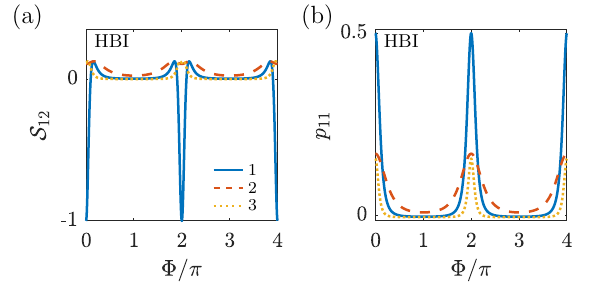}
\end{center}
\caption{(a,b)~Phase-dependence of~${\cal S}_{12}$ and~$p_{11}$ for the pairs~($R_1$, $R_2$) indicated on Fig.3(b-d).
}
\label{figSM}
\end{figure}
\clearpage

\section{Current cross-correlator and splitting probability in a line-shaped NSN~device}
\begin{figure}[h]
\begin{center}
\includegraphics[width=.95\linewidth]{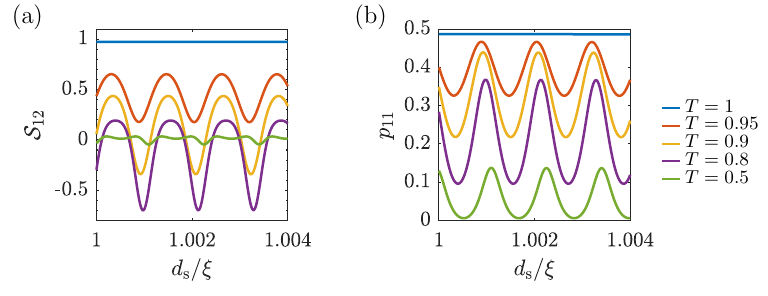}
\end{center}
\caption{(a,b)~Oscillations of ${\cal S}_{12}$ and $p_{11}$ with superconductor length for a set of S/N interface transparencies for the case of~$d_{\text{s}}\approx\xi$. Note that due to oscillations ${\cal S}_{12}$ becomes on the average negative at $T=0.8$.
}
\label{fig_osc}
\end{figure}

\begin{figure}[h]
\begin{center}
\includegraphics[width=.85\linewidth]{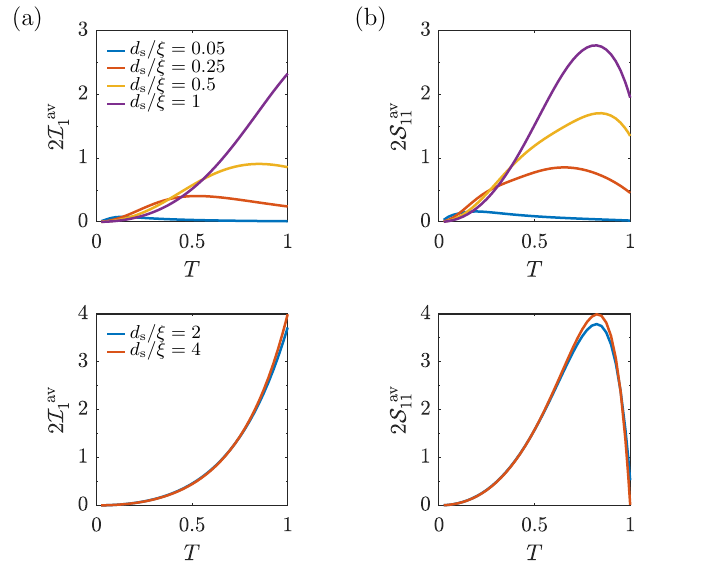}
\end{center}
\caption{(a,b)~This figure shows the values of $2{\cal I}_{1}$ and $2{\cal S}_{11}$ obtained after averaging over one period of oscillations. Our results perfectly reproduce the results of~\cite{Freyn2010}, see Fig.3, left and central panels.}
\label{fig_av}
\end{figure}

\begin{figure}[h]
\begin{center}
\includegraphics[width=.85\linewidth]{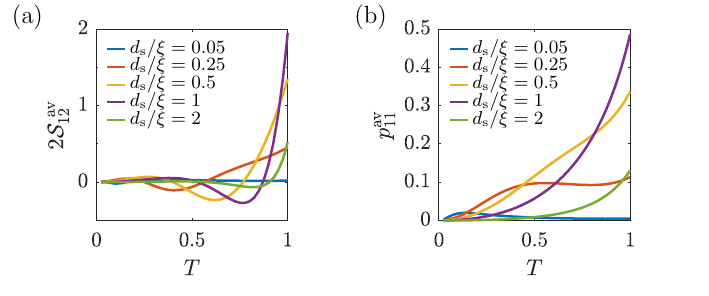}
\end{center}
\caption{(a,b)~This figure shows the values of $2{\cal S}_{12}$ and $p_{11}$ obtained after averaging over one period of oscillations. Panel~(a) perfectly reproduces the results of~\cite{Freyn2010}, see Fig.3, right panel.}
\label{fig_av}
\end{figure}

\begin{figure}[h]
\begin{center}
\includegraphics[width=.85\linewidth]{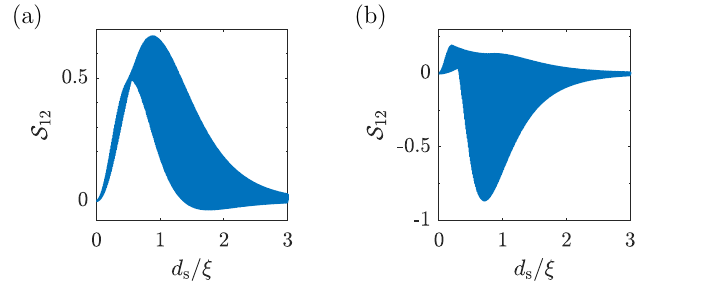}
\end{center}
\caption{Oscillations in ${\cal S}_{12}$ at (a)~$T=0.95$ and (b)~$T=0.75$.}
\label{fig_S12_show_neg}
\end{figure}
\end{widetext}
\end{document}